\documentclass[twocolumn]{aastex631}

\shorttitle{USS: Photometric Re-calibration with the BEST Database}
\shortauthors{Li et al.}
\graphicspath{{./}{figures/}}

\usepackage{subfigure}
\usepackage{amsmath}
\usepackage{CJKutf8}

\begin{document}
\begin{CJK}{UTF8}{gbsn}

\title{The S-PLUS Ultra-Short Survey: Photometric Re-calibration with the BEst STar Database}

\correspondingauthor{Kai Xiao, \\ Yang Huang, Haibo Yuan, Yanke Tang}
\email{xiaokai@ucas.ac.cn; huangyang@ucas.ac.cn; \\ yuanhb@bnu.edu.cn; tyk@dzu.edu.cn}

\author[0009-0007-8288-5656]{Xiaolu Li}
\affiliation{College of Physics and Electronic information, Dezhou University, Dezhou 253023, Peopleʼs Republic of China}
\affiliation{International Centre of Supernovae, Yunnan Key Laboratory, Kunming 650216, People's Republic of China}

\author[0000-0001-8424-1079]{Kai Xiao}
\affiliation{School of Astronomy and Space Science, University of Chinese Academy of Sciences, Beijing 100049, People's Republic of China}
\affiliation{Institute for Frontiers in Astronomy and Astrophysics, Beijing Normal University, Beijing, 102206, China}

\author[0000-0003-3250-2876]{Yang Huang}
\affiliation{School of Astronomy and Space Science, University of Chinese Academy of Sciences, Beijing 100049, People's Republic of China}
\affiliation{CAS Key Lab of Optical Astronomy, National Astronomical Observatories, Chinese Academy of Sciences, Beijing 100012, People's Republic of China}

\author[0000-0003-2471-2363]{Haibo Yuan}
\affiliation{Institute for Frontiers in Astronomy and Astrophysics, Beijing Normal University, Beijing, 102206, China}
\affiliation{School of Physics and Astronomy, Beijing Normal University No.19, Xinjiekouwai St, Haidian District, Beijing, 100875, China}

\author[0000-0003-4207-1694]{Yanke Tang}
\affiliation{College of Physics and Electronic information, Dezhou University, Dezhou 253023, Peopleʼs Republic of China}
\affiliation{International Centre of Supernovae, Yunnan Key Laboratory, Kunming 650216, People's Republic of China}

\author[0000-0003-4573-6233]{Timothy C. Beers}
\affiliation{Department of Physics and Astronomy and JINA Center for the Evolution of the Elements (JINA-CEE), University of Notre Dame, Notre Dame, IN 46556, USA}

\author[0000-0002-1259-0517]{Bowen Huang}
\affiliation{Institute for Frontiers in Astronomy and Astrophysics, Beijing Normal University, Beijing, 102206, China}
\affiliation{School of Physics and Astronomy, Beijing Normal University No.19, Xinjiekouwai St, Haidian District, Beijing, 100875, China}

\author[0009-0003-1069-1482]{Mingyang Ma}
\affiliation{Institute for Frontiers in Astronomy and Astrophysics, Beijing Normal University, Beijing, 102206, China}
\affiliation{School of Physics and Astronomy, Beijing Normal University No.19, Xinjiekouwai St, Haidian District, Beijing, 100875, China}

\author[0000-0003-3537-4849]{Pedro K. Humire}
\affil{Departamento de Astronomia, Instituto de Astronomia, Geofísica e Ciências Atmosféricas da USP, Cidade Universitária, 05508-900 São Paulo, SP, Brazil} 

\author[0000-0002-5045-9675]{Alvaro Alvarez-Candal}
\affil{Instituto de Astrofísica de Andalucía, CSIC, Apt 3004, 18080 Granada, Spain}

\author{Federico Sestito}
\affil{Centre for Astrophysics Research, Department of Physics, Astronomy and Mathematics, University of Hertfordshire, Hatfield, AL10 9AB, UK}

\author[0000-0002-9308-587X]{Ning Gai}
\affiliation{College of Physics and Electronic information, Dezhou University, Dezhou 253023, Peopleʼs Republic of China}
\affiliation{International Centre of Supernovae, Yunnan Key Laboratory, Kunming 650216, People's Republic of China}

\author{Yongna Mao}
\affiliation{CAS Key Lab of Optical Astronomy, National Astronomical Observatories, Chinese Academy of Sciences, Beijing 100012, People's Republic of China}

\author[0009-0007-5610-6495]{Hongrui Gu}
\affiliation{CAS Key Lab of Optical Astronomy, National Astronomical Observatories, Chinese Academy of Sciences, Beijing 100012, People's Republic of China}
\affiliation{School of Astronomy and Space Science, University of Chinese Academy of Sciences, Beijing 100049, People's Republic of China}

\author[0000-0002-4683-5500]{Zhenzhao Tao}
\affiliation{College of Computer and Information, Dezhou University, Dezhou 253023, China}
\affiliation{Institute for Astronomical Science, Dezhou University, Dezhou 253023, China}

\author[0000-0002-9824-0461]{Lin Yang}
\affiliation{Department of Cyber Security, Beijing Electronic Science and Technology Institute, Beijing, 100070, China}

\author[0000-0003-3535-504X]{Shuai Xu}
\affiliation{Institute for Frontiers in Astronomy and Astrophysics, Beijing Normal University, Beijing, 102206, China}
\affiliation{School of Physics and Astronomy, Beijing Normal University No.19, Xinjiekouwai St, Haidian District, Beijing, 100875, China}

\author{Rong Hu}
\affiliation{Department of Physics, Beijing Technology and Business University, Beijing 100048, People's Republic of China}

\journalinfo{Accepted by ApJS on January 6, 2025}

\begin{abstract}
We present an independent validation and comprehensive re-calibration of S-PLUS Ultra-Short Survey (USS) DR1 12-band photometry using about 30,000--70,000 standard stars from the BEst STar (BEST) database. We identify spatial variation of zero-point offsets, up to 30--40\,mmag for blue filters ($u$, $J0378$, $J0395$) and 10\,mmag for others, predominantly due to the higher uncertainties of the technique employed in the original USS calibration. Moreover, we detect large- and medium-scale CCD position-dependent systematic errors, up to 50\,mmag, primarily caused by different aperture and flat-field corrections. We then re-calibrate the USS DR1 photometry by correcting the systematic shifts for each tile using second-order two-dimensional polynomial fitting combined with a numerical stellar flat-field correction method. The re-calibrated results from the XPSP and the SCR standards are consistent within 6\,mmag in the USS zero-points, demonstrating both the typical precision of re-calibrated USS photometry and a sixfold improvement in USS zero-point precision. Further validation using SDSS and Pan-STARRS1, as well as LAMOST DR10 and Gaia photometry, also confirms this precision for the re-calibrated USS photometry.  Our results clearly demonstrate the capability and the efficiency of the BEST database in improving calibration precision to the milli-magnitude level for wide-field photometric surveys. The re-calibrated USS DR1 photometry is publicly available (\href{https://nadc.china-vo.org/res/r101504/}{doi: 10.12149/101503}).
\end{abstract}

\keywords{Stellar photometry, Astronomy data analysis, Calibration}

\vskip 2cm
\section{Introduction} 
\label{sec:intro}
The current era of astronomy is characterized by the flourishing of wide-field photometric survey projects, with a continuous influx of photometric data significantly impacting various fields within the discipline. Ensuring the consistency of flux measurements between widely separated targets under varying observing conditions, across different detector positions, and at different observing times is crucial for the success of these projects. 
Achieving high-precision photometric calibration is essential to conduct high-precision scientific research because the precision of photometric calibration limits the detection accuracy in astronomical measurements.

Achieving milli-magnitude level photometric calibration in ground-based photometric surveys presents a significant challenge, due to the complexity of accounting for systematic errors introduced by instrumental effects and the Earth's atmospheric influence \citep{2006ApJ...646.1436S}. 
Historically, the limited number and precision of photometric standard stars have made it difficult to accurately measure and correct these errors. However, the recent release of extensive stellar atmospheric parameters from large-scale spectroscopic surveys such as LAMOST
\citep{2012RAA....12.1197C,2012RAA....12..735D,2012RAA....12..723Z,2014IAUS..298..310L}, along with Gaia photometry and BP/RP (BP and RP are the abbreviations for Blue Photometer and Red Photometer, respectively; BP/RP is often shortened as XP) spectra \citep{2023A&A...674A...1G}, offers an opportunity to establish a large array of high-precision photometric standard stars. This advancement enables more accurate measurement and correction of the complex systematic errors.

In recent years, a series of photometric calibration methods have been proposed \citep{2022ApJS..259...26H,2022SSPMA..52B9503H}, including both ``hardware/observation-driven'' and ``software/physics-driven'' methods. Among them, the stellar color regression (SCR) method and the Gaia XP spectra based synthetic photometry (XPSP) method have achieved outstanding results. The SCR method, first proposed by \cite{2015ApJ...799..133Y}, predicts the intrinsic colors of stars using a few physical quantities. For example, the stellar atmospheric parameters can be used to predict the intrinsic colors. Applied to the Sloan Digital Sky Survey \citep[SDSS;][]{2000AJ....120.1579Y} Stripe 82 \citep{2007AJ....134..973I}, the SCR method achieved a three-fold improvement in color and magnitude accuracy, with precision levels of 2--5\,mmag \citep{2015ApJ...799..133Y,2022ApJS..259...26H}. Application of this method to Gaia Data Release 2 \citep{2018A&A...616A...1G} and Early Data Release 3 \citep[EDR3, ][]{2021A&A...649A...1G,2021A&A...650C...3G} resulted in the correction of systematic errors to a precision of 1\,mmag \citep{2021ApJ...909...48N,2021ApJ...908L..14N,2021ApJ...908L..24Y}. The SCR method was also used to re-calibrate the SkyMapper Southern Survey \citep[SMSS;][]{2018PASA...35...10W} Data Release 2, identifying significant zero-point offsets in the $u$- and $v$-bands \citep{2021ApJ...907...68H}. Furthermore, the method corrected spatial- and magnitude-dependent systematic errors in Pan-STARRS1 \citep[PS1;][]{2012ApJ...750...99T} Data Release 1, achieving of 1--2\,mmag precision at a spatial resolution of $14^{\prime\prime}$ \citep{2022AJ....163..185X,2023ApJS..268...53X}, and was also applied to Stellar Abundance and Galactic Evolution Survey \citep[SAGES;][]{Zheng18,Zheng19,2023ApJS..268....9F} photometry, achieving precisions of 1--2\,mmag for the $gri$-bands \citep{xiao} and $\le 5$\,mmag for the $uv$-bands.

In June 2022, Gaia DR3 released XP spectra for approximately 220 million sources \citep{2021A&A...652A..86C,2023A&A...674A...1G}, mostly with magnitudes $G<17.65$ and covering wavelengths from 336 to 1020\,nm, which were calibrated both internally \citep{2021A&A...652A..86C,2023A&A...674A...2D} and externally \citep{2023A&A...674A...3M}. 
Based on the Gaia XP spectra, \cite{2023A&A...674A..33G} proposed the XPSP method. However, Gaia XP spectra exhibit systematic errors related to magnitude, color, and extinction; particularly below 400\,nm \citep{2023A&A...674A...3M,2024ApJS..271...13H}. Recently, \cite{2024ApJS..271...13H} performed comprehensive corrections on the Gaia XP spectra. The \textit{corrected} Gaia XP spectra were used to improve the XPSP method \citep[][]{2023ApJS..269...58X}, which no longer relies on 343 spectra coefficients and can directly derive multi-band magnitudes from the corrected spectra. This method achieves higher accuracy in constructing photometric standard stars. The improved XPSP method was applied to the photometric re-calibration of J-PLUS DR3 data \citep{2024A&A...683A..29L}, achieving zero-point precision of 1--5\,mmag -- a two-fold improvement \citep{2023ApJS..269...58X}. 

More recently, using the SCR and improved XPSP methods, Xiao et al. (in prep.) created the BEst STar (BEST) database, which includes over 200 million high-precision photometric standard stars \footnote{Here, the photometric standard stars are a series of stars whose uniformity and accuracy of magnitudes have been meticulously ensured across the 12 passbands of the S-PLUS photometric system.}. This database covers more than ten photometric systems, such as the Gaia \citep{2021A&A...649A...1G,2021A&A...650C...3G}, Landolt \citep{2013AJ....146...88C,2016AJ....152...91C}, J-PLUS \citep{2019A&A...622A.176C}, S-PLUS \citep{2019MNRAS.489..241M}, J-PAS \citep{2014arXiv1403.5237B}, SDSS \citep{2000AJ....120.1579Y}, PS1 \citep{2012ApJ...750...99T}, CSST \citep{2018cosp...42E3821Z}, LSST \citep{2019ApJ...873..111I}, SkyMapper \citep{2018PASA...35...10W}, and SiTian project \citep{2021AnABC..93..628L} systems, and spans hundreds of bands, providing coverage for stars brighter than 17.65\,mag across the entire sky. In the process, we first predicted the XPSP standard magnitudes across those photometric systems based on the \textit{corrected} Gaia XP spectra by \cite{2024ApJS..271...13H} and the total transmission function of each photometric band using an improved XPSP method. For bands slightly beyond the wavelength coverage of Gaia XP spectra, such as the SDSS $u$-band, we used a linear fit to extrapolate the spectra, following the approach described by \cite{2023ApJS..269...58X}. Following that, using the extensive stellar atmospheric parameters provided by LAMOST \citep{2012RAA....12.1197C,2012RAA....12..735D,2012RAA....12..723Z,2014IAUS..298..310L}, GALAH \citep{2015MNRAS.449.2604D}, and others, as well as the XPSP magnitudes, we employed the SCR method to derive the SCR standard magnitudes for stars with known stellar atmosphere parameters. There is good consistency between the SCR and the XPSP standard photometry, as noted by Xiao et al.  (in prep.). Below we present a detailed comparison of the consistency between the standard stars constructed using these two methods across the S-PLUS ultra-short survey (USS) photometric filters.

As a sub-survey of the S-PLUS survey, the USS employs a 12-filter system \citep{2024arXiv240705004P}, comprising seven narrow/medium-band filters and five broad-band filters. This imaging survey spans the same sky area as the overall S-PLUS Main Survey, but with significantly shorter exposure times of 3--20\,s (1/40th to 1/33th of those of the S-PLUS Main Survey). The primary goal of the USS is to uncover bright, extremely metal-poor (EMP; $-4.0 \le {\rm [Fe/H]} < -3.0$) and ultra metal-poor (UMP; [Fe/H] $< -4.0$) stars. As \cite{2024ApJ...968L..24X} highlighted, the accuracy of photometric data and the sensitivity of the photometric bands are both crucial in photometry-based stellar metallicity measurement. To achieve this objective, high-precision photometric calibration of the USS DR1 data is essential. 

In this study, we perform a photometric re-calibration of the USS DR1 data aiming to achieve uniform photometry with an accuracy at the milli-magnitude level using the BEST database.
The structure of this paper is as follows. We present the dataset used in this work in Section \ref{sec:data}. In Section\,\ref{sec:vali}, we provide a detailed description of the independent calibration validation of the USS DR1 data, followed by the correction of the systematic errors in Section\,\ref{sec:correct}. Section\,\ref{sec:s5} addresses the zero-point precision of the photometric re-calibration. A discussion is presented in Section\,\ref{sec:dis}, followed by a summary and conclusions provided in Section\,\ref{sec:conclusion}.

\section{Data} \label{sec:data}
\subsection{USS DR1} \label{sec:uss}
The USS DR1 includes 163 tiles along the Celestial Equator for all bands, with saturation magnitudes of about $10$th magnitude (in the broad-band SDSS-like filters), with each tile measuring 2 square degrees \citep{2024arXiv240705004P}. It is important to note that USS continues observing even when not all observations are conducted under optimal photometric conditions, such as when cirrus clouds are present.
As detailed in \cite{2024arXiv240705004P}, the USS provides aperture instrumental photometry obtained using SExtractor \citep{1996A&AS..117..393B}. Aperture corrections were performed based on the 3-arcsec aperture magnitudes by measuring the magnitudes in 32 concentric apertures centered around each point source, averaging the magnitude changes in larger apertures until convergence, considering sources with signal-to-noise ratios (SNR) between 30 and 1000.
The result of the aperture correction is constant for each tile, as shown in Figure\,\ref{Fig:A0}. For photometric calibration, \cite{2024arXiv240705004P} first used stellar spectral energy distributions (SED) fitting to convert PS1 standard stars from the ATLAS all-sky stellar reference catalog \citep[ATLAS Refcat2;][]{2018ApJ...867..105T} to the USS system for the $g$-, $J0515$-, $r$-, $J0660$-, $i$-, $J0861$-, and $z$-band photometry, then used the stellar locus (SL; \citealt{2022MNRAS.511.4590A}) method to calibrate the $u$-, $J0378$-, $J0395$-, $J0410$-, and $J0430$-bands based on the calibrated $g$ and $i$ magnitudes (e.g., $u-g$ versus $g-i$), then finally performed SED-based calibration for all 12 bands.

The final catalog of USS DR1 includes photometrically calibrated magnitudes measured in 3-arcsec (labelled \texttt{APER\_3}) and 6-arcsec (\texttt{APER\_6}) diameter apertures, aperture corrected \texttt{APER\_3} magnitudes (\texttt{PStotal}), and 
stellar- profile information, such as the normalized full width at half maximum (FWHM; labelled \texttt{FWHM\_n}) and ellipticity (\texttt{ELLIPTICITY}). All USS DR1 data is publicly available through the splus.cloud service \footnote{\url{https://splus.cloud/}}. The USS magnitudes mentioned in this paper refer to the \texttt{PStotal} magnitudes.

We stress that, if the assumption that a 3-arcsec aperture is the best aperture is flawed, the aperture correction of USS DR1 may be sub-optimal. Moreover, systematic errors from external reference catalogs (e.g., ATLAS Refcat2) might propagate into the USS data, and the SL method's different treatment of stellar metallicity may introduce metallicity-dependent systematics in the blue filters. Additionally, the spatial inhomogeneity of image quality and astrometric centering, as well as a different flat-field correction, could cause CCD position-dependent systematic errors in USS DR1.

\begin{figure}[ht!] \centering
\resizebox{\hsize}{!}{\includegraphics{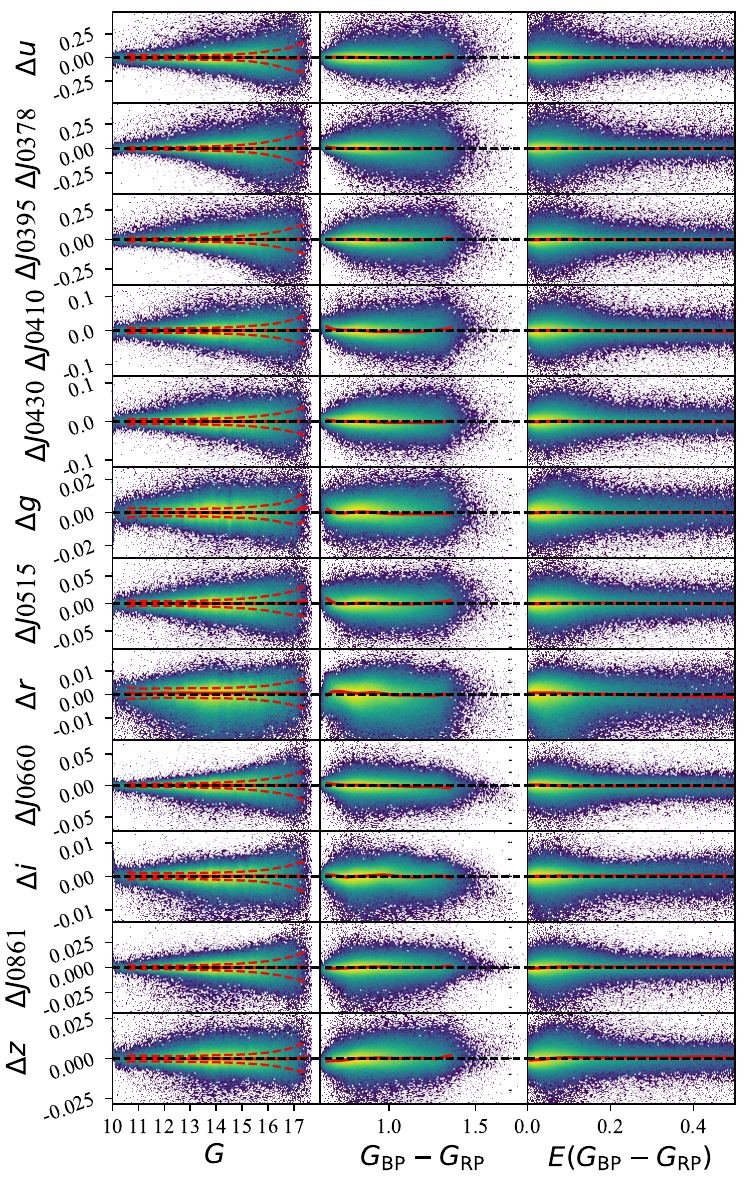}}
\caption{{\small Magnitude offsets between the XPSP and the SCR standard stars, as functions of Gaia $G$, $G_{\rm BP}-G_{\rm RP}$, and $E(G_{\rm BP}-G_{\rm RP})$, for all 12 bands. The colors represent the density of the points. For each panel, zero residuals and median values of magnitude offsets are denoted by the black-dashed lines and red curves, respectively. The red dashed curves in the left panel represent the standard deviations of the points, estimated using Gaussian fitting with a running width of 0.5\,mag and a running step of 0.2\,mag.}}
\label{Fig:f1}
\end{figure}

\begin{table}[ht!] 
\centering
\caption{Standard Deviations of the Gaia $G$ Magnitude Differences between the SCR and XPSP Standard Stars}
\begin{tabular}{c|c|c|c|c}
\hline
\hline
Filter   & $G\sim 11$ & $G\sim 13$ & $G\sim 15$   & $G\sim 17$\\ \hline
$J$0378    &     0.0142  &  0.0174    &     0.0372  & 0.1329 \\
$J$0395    &     0.0132  &  0.0155    &     0.0306  & 0.0971 \\
$J$0410    &     0.0060  &  0.0068    &     0.0119  & 0.0322 \\
$J$0430    &     0.0061  &  0.0067    &     0.0111  & 0.0281 \\
$J$0515    &     0.0041  &  0.0047    &     0.0088  & 0.0219 \\
$J$0660    &     0.0024  &  0.0032    &     0.0063  & 0.0170 \\
$J$0861    &     0.0017  &  0.0022    &     0.0041  & 0.0111 \\
$u$      &     0.0206  &  0.0213    &     0.0362  & 0.1244 \\
$g$      &     0.0022  &  0.0023    &     0.0032  & 0.0074 \\
$r$      &     0.0020  &  0.0020    &     0.0025  & 0.0051 \\
$i$      &     0.0009  &  0.0009    &     0.0013  & 0.0035 \\
$z$      &     0.0017  &  0.0019    &     0.0028  & 0.0069 \\ \hline 
\end{tabular}
\label{tab1}
\end{table}

\begin{figure*}[ht!] \centering
\resizebox{\hsize}{!}{\includegraphics{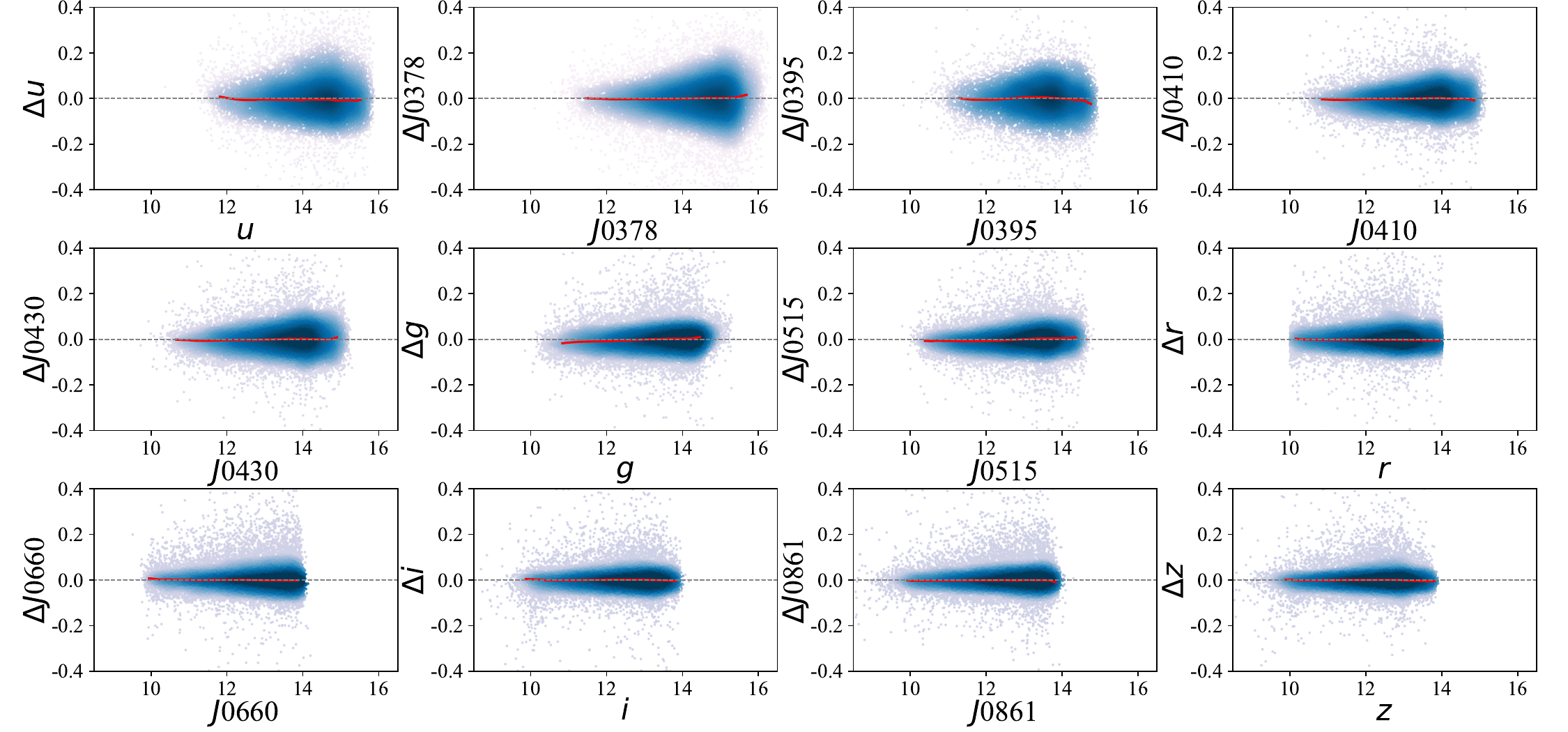}}
\caption{{\small 
Variations of the magnitude offsets, as a function of USS magnitude, for all 12 bands. For each panel, the colors represent the logarithm of the number density, calculated using Gaussian kernel density estimation. The red and gray lines denote the median values and zero level of the magnitude offsets, respectively.}}
\label{Fig:f2}
\end{figure*}

\begin{figure*}[ht!] \centering
\resizebox{\hsize}{!}{\includegraphics{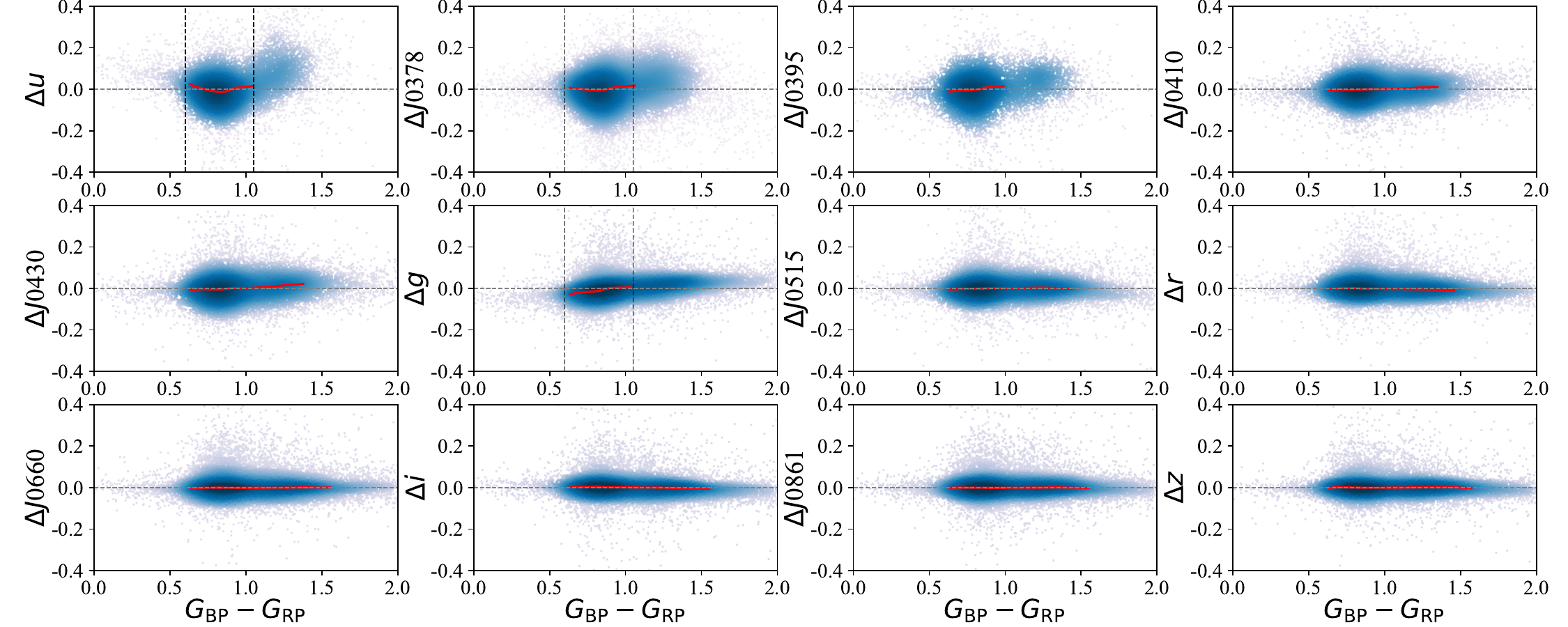}}
\caption{{\small 
Same as Figure\,\ref{Fig:f2}, but for color $G_{\rm BP}-G_{\rm RP}$. For the $u$-, $J0378$-, $J0395$-, and $g$-bands; two color cuts are denoted by black-dashed lines. }}
\label{Fig:f3}
\end{figure*}

\begin{figure*}[ht!] \centering
\resizebox{\hsize}{!}{\includegraphics{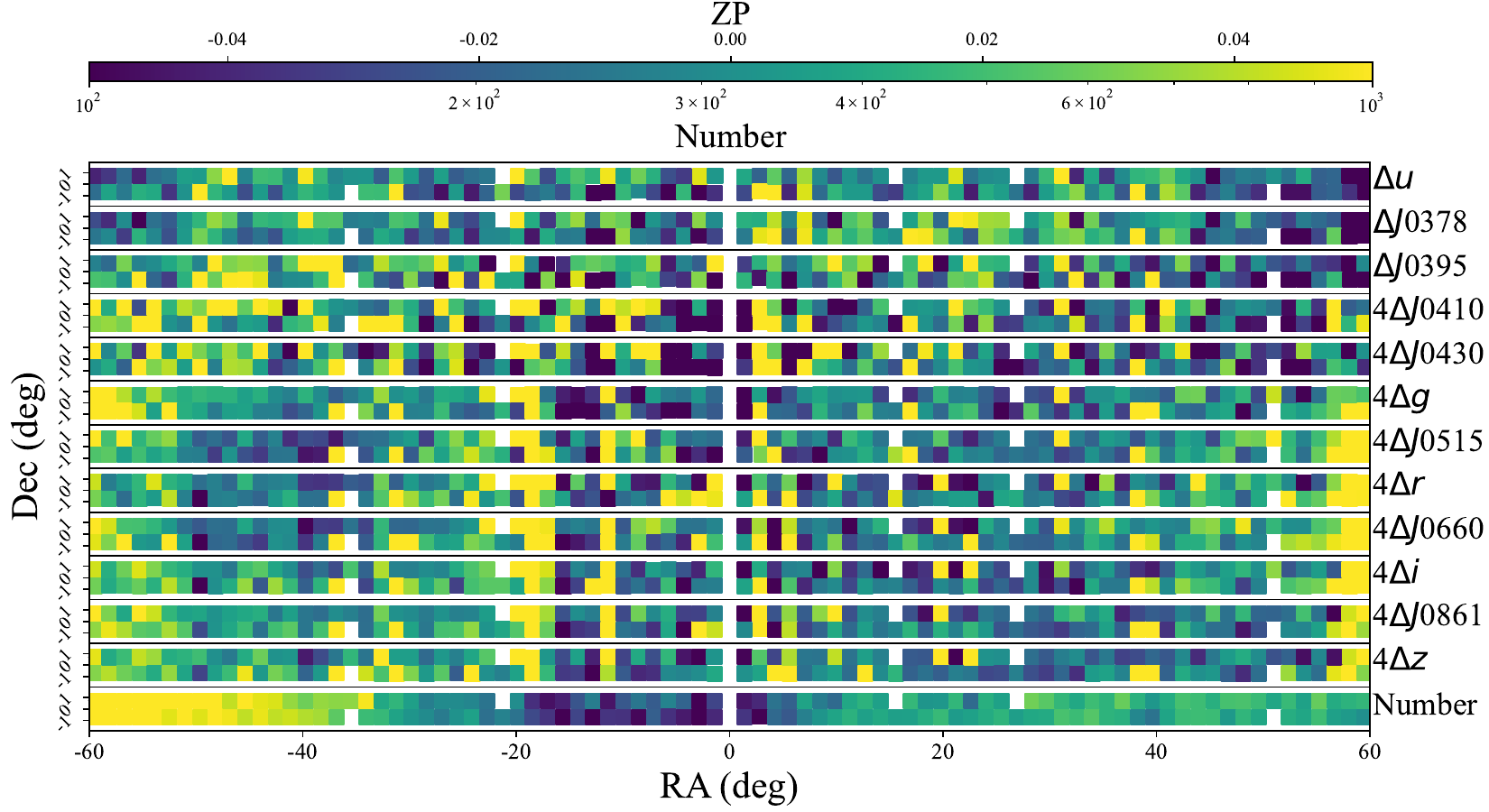}}
\caption{{\small 
From top to bottom, the first 12 rows show spatial variations of the difference between the XPSP standards and USS zero-points in each tile. The bottom row shows the distributions of the listed numbers of each tile in the $r$-band. 
To clearly illustrate the spatial structure, the zero-point differences in the $J0410$-, $J0430$-, $g$-, $J0515$-, $r$-, $J0660$-, $i$-, $J0861$-, and $z$-bands are multiplied by four. 
The labels are marked in each panel, and a color bar is shown on the top.}}
\label{Fig:f4}
\end{figure*}

\subsection{BEST Standards in the USS System} \label{sec:best}
The BEST database (Xiao et al., in prep.) provides (1) a 12-band standard star catalog in the USS system, encompassing approximately 5 million SCR standard stars predicted using the SCR method based on spectroscopic data from the LAMOST \citep{2012RAA....12.1197C,2012RAA....12..735D,2012RAA....12..723Z,2014IAUS..298..310L} DR10 (\href{https://www.lamost.org/dr10/}{https://www.lamost.org/dr10/}) and corrected Gaia EDR3 photometric data \citep{2021ApJ...908L..24Y}, and over (2) 200 million XPSP standard stars derived using an improved XPSP method \citep{2023ApJS..269...58X} based on ``corrected'' Gaia XP spectra \citep{2024ApJS..271...13H}. For reddening correction in the SCR method, we avoid using the dust reddening map by \cite{1998ApJ...500..525S}, as it has proven to be unreliable at low Galactic latitudes, and exhibits spatially-dependent systematic errors \citep{2022ApJS..260...17S}. Instead, we employ the star-pair method \citep{2013MNRAS.430.2188Y,2020ApJ...905L..20R} to determine the values of $E(G_{\rm BP}-G_{\rm RP})$.

The XPSP standard stars are distributed across the entire sky, with magnitudes ranging from about 10th to 17.65\,mag. The SCR standard stars are main-sequence stars (${\log g}>-3.4\times 10^{-4}\times T_{\rm eff}+5.8$), confined to the sky region north of declination $-10^{\circ}$, and comply with the following constraints: $4500\le T_{\rm eff} \le 6500$\,K and $\rm [Fe/H] \ge -1$ for robust fitting in the SCR method, \texttt{phot}\_\texttt{bp}\_\texttt{rp}\_\texttt{excess}\_\texttt{factor} $<$ $1.3+0.06\times(G_{\rm BP}-G_{\rm RP})^2$ to avoid poor Gaia $G_{\rm BP}/G_{\rm RP}$ photometry, and a SNR for the LAMOST $g$-band spectra greater than $20$. 

To evaluate the consistency between the magnitudes of the SCR and the XPSP standard stars, Figure\,\ref{Fig:f1} presents their difference as functions of Gaia $G$ magnitude, Gaia color $G_{\rm BP}-G_{\rm RP}$, and extinction $E(G_{\rm BP}-G_{\rm RP})$. Due to the low-extinction ($E(B-V)\le 0.068$\,mag; \citealt{2024arXiv240705004P}) of stars in the USS DR1, only stars with extinction $E(G_{\rm BP}-G_{\rm RP})\le 0.5$ are considered here. As expected, no dependence on magnitude, color, or extinction is found. At the bright end, the standard deviation is smallest, but as the stellar brightness decreases, it first increases slowly and then rapidly. For example, in the $i$-band, the scatter is only 0.9\,mmag for $G\sim 11$, increases slightly to 1.3\,mmag at $G=15$, and then quickly grows to 3.5\,mmag at the faint end of $G\sim 17$. Table\,\ref{tab1} presents the scatter values of the magnitude differences between the SCR and the XPSP standards at four specific magnitudes for all 12 bands.

\subsection{Calibration Stars} \label{sec:uss}
We combine the USS DR1 photometric data with the BEST catalog using a adopted cross-matching radius of $1''$. Then, we select standard stars as the calibration samples with the following constraints:
\begin{enumerate}
\item[(a)] The magnitude in the $r$-band is greater than 10\,mag to avoid saturation, following \cite{2024arXiv240705004P}.
\item[(b)] Photometric error\{$u$, $J0395$, $J0410$, $J0430$\} $<$ 0.05\,mag, error\{$J0378$\} $<$ 0.08\,mag, error\{$J0515$\} $<$ 0.03\,mag, error\{$g$, $J0660$, $J0861$\} $<$ 0.02\,mag, and error\{$r$, $i$, $z$\} $<$ 0.01\,mag to maintain a balance between good SNR and a sufficient number of photometric standard stars for each band.
\item[(c)] \texttt{SEX\_FLAGS} $=$ 0 to avoid bad USS DR1 photometry.
\end{enumerate}
Finally, 26118, 44546, 31395, 31891, 37732, 56556, 40081, 35303, 81081, 46678, 60970, and 38423 calibration stars are selected in the $u$-, $J0378$-, $J0395$-, $J0410$-, $J0430$-, $g$-, $J0515$-, $r$-, $J0660$-, $i$-, $J0861$-, and $z$-bands, respectively.

\begin{figure*}[ht!] \centering
\includegraphics[width=16.8cm]{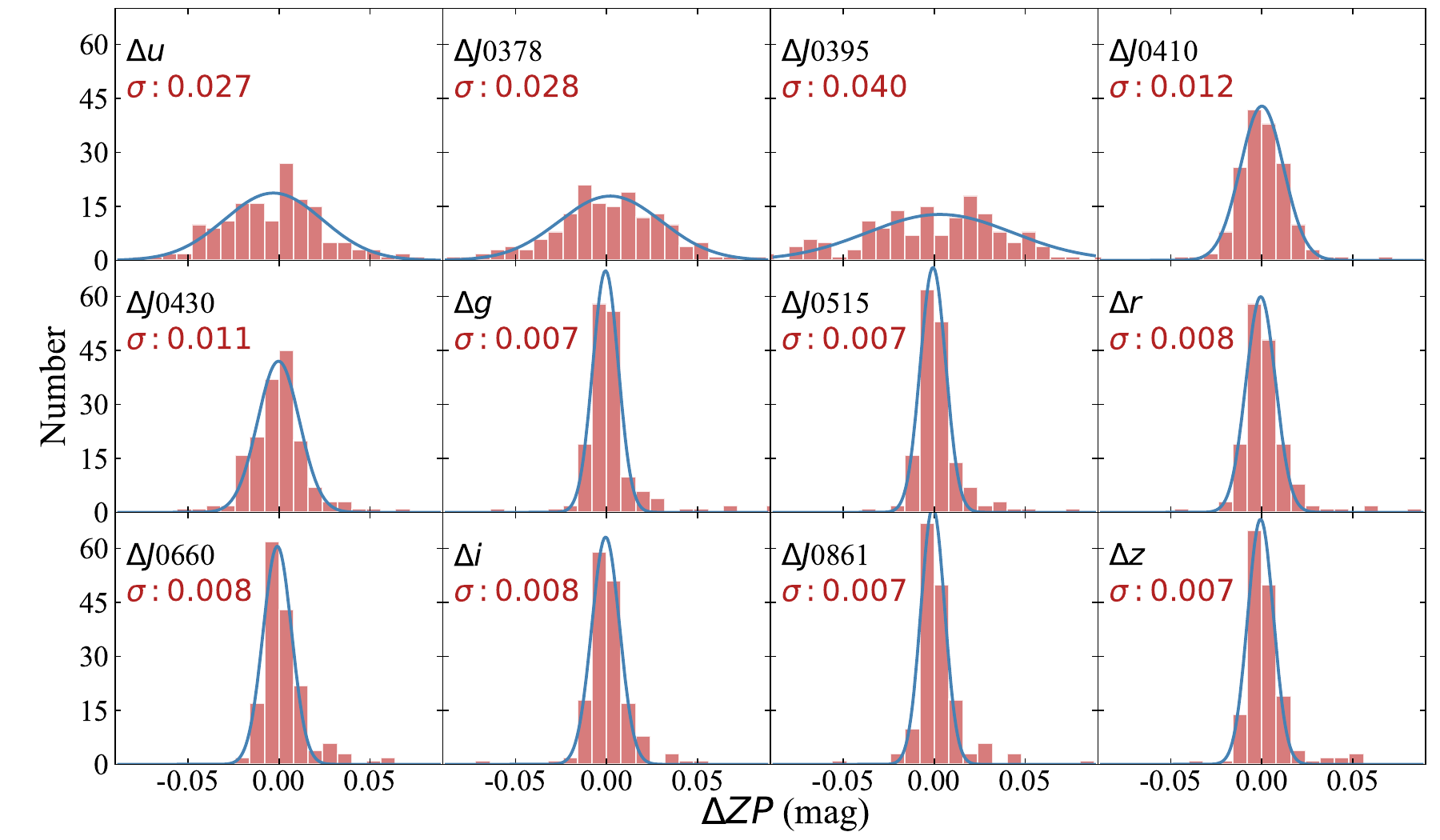}
\caption{{\small Histograms of the differences in the zero-points between the XPSP standards and USS photometry. The bands are marked on the top-left corner of each panel. The Gaussian fitting results are plotted as blue lines. The standard deviations using Gaussian fitting are marked in each panel in red.}}
\label{Fig:f5}
\end{figure*}

\begin{figure*}[ht!] \centering
\resizebox{\hsize}{!}{\includegraphics{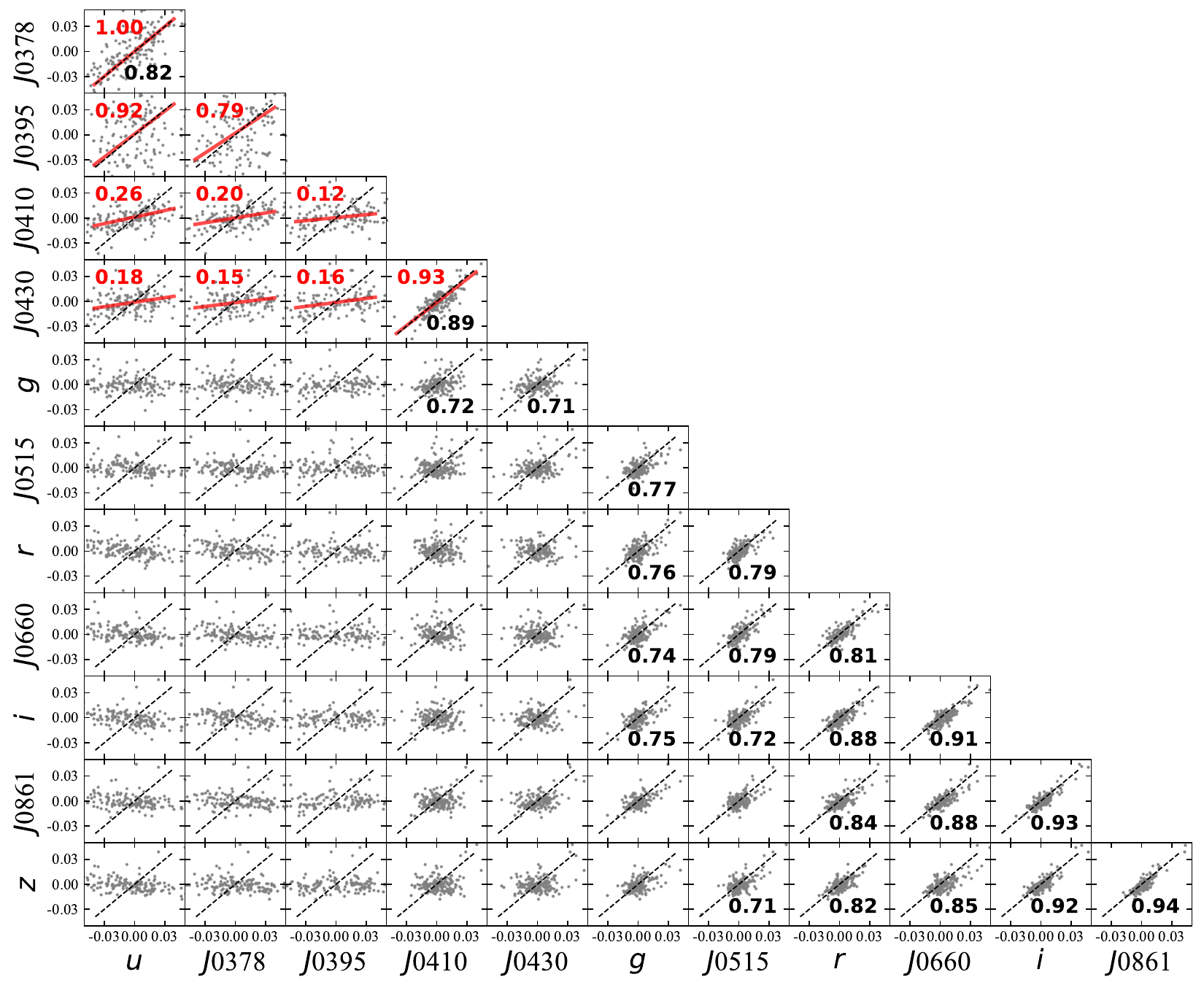}}
\caption{{\small Correlation plots of the zero-point offsets between the BEST standards and USS photometry for each of the two bands. For each panel, the correlation coefficient is displayed in the bottom-right corner only when it exceeds 0.7. The black-dashed lines denote $y=x$ in each panel. The red lines represent the results of linear fitting;  the slopes are also marked in red. }}
\label{Fig:f6}
\end{figure*}

\vskip 1cm
\section{Validation of the USS DR1 Photometric Calibration} \label{sec:vali}
As previously noted by \cite{2022SSPMA..52B9503H}, systematic errors in photometric data can be mathematically expressed as functions of time, magnitude, color, and the star's position on the detector. For example, short-term variations in the Earth's atmosphere cause zero-point shifts in each observation, the non-linearity effects of detector (e.g., CCDs) introduces magnitude-dependent errors, the color effects of Earth's atmospheric and differences in photometric system could lead to color-related systematics, and different flat-field corrections result in errors depending on the detector position ($X$ and $Y$). To precisely measure and describe possible potential systematic errors in USS DR1 data, we first validate dependencies of the magnitude difference between the BEST standards and the USS on magnitude, color, and detector position.

\subsection{Dependence of the Magnitude Offsets on USS Magnitude and Gaia $G_{\rm BP}-G_{\rm RP}$ Color} \label{sec:mag_color}
The magnitude offsets between the XPSP standards and USS, as functions of the USS DR1 magnitude and $G_{\rm BP}-G_{\rm RP}$, are shown in Figure\,\ref{Fig:f2} and Figure\,\ref{Fig:f3}, respectively. No dependence on USS magnitude is found for all bands, suggesting that the USS detector exhibits a high level of linearity. However, for the $u$-, $J0378$-, and $g$- bands, a slight dependence on $G_{\rm BP}-G_{\rm RP}$ color is found, especially when the color is greater than 1.05 or less than 0.6. We attribute this effect to measurement errors in the response curve of the USS $u$-, $J0378$-, and $g$-bands and/or Gaia DR3 XP spectra. Additionally, we must consider the potential influence of extrapolating the XP spectra beyond the $u$-band.

For the calibration of the $u$-, $J0378$-, and $g$-bands, we selectively choose stars within the specific $G_{\rm BP}-G_{\rm RP}$ range of 0.6 to 1.05; less than 3\%  of the stars fall outside this range.

\subsection{Spatially Dependent Systematics in Zero-points} \label{sec:image}

\begin{figure*}[ht!] \centering
\resizebox{\hsize}{!}{\includegraphics{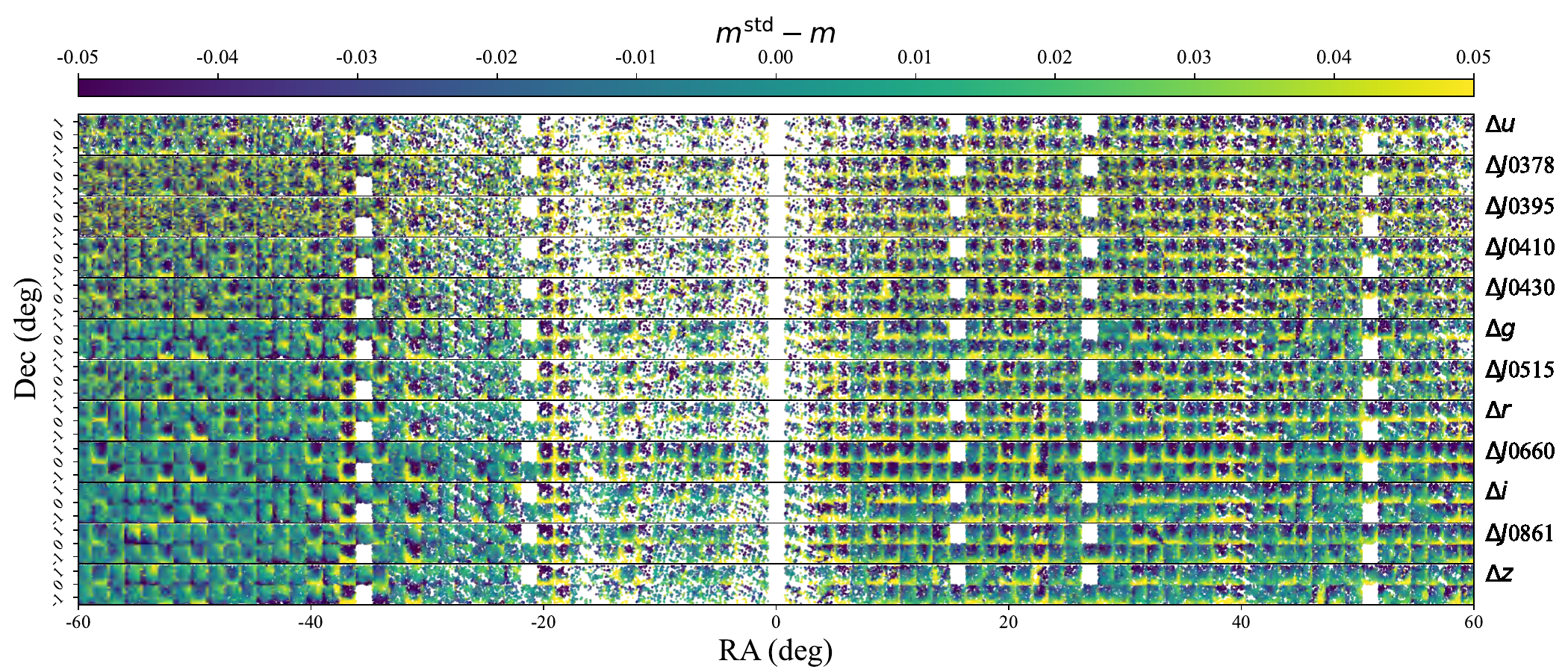}}
\caption{{\small 
Same as Figure\,\ref{Fig:f4}, but for magnitude offsets between the XPSP standards and the USS photometry after subtraction of the zero-points for each tile. The bands are marked on the right, and a color bar is shown on the top. The tire-track like structures observed in the figure, extending roughly from $\rm RA=-30^{\circ}$ to 0 and related to Gaia's scanning law, are mainly caused by the spatial non-uniformity of the numbers of Gaia XP spectra. Similar structures are also observed in the bottom-right panel of Figure\,4 in \cite{2023ApJS..268...53X}. 
}}
\label{Fig:f7}
\end{figure*}

\begin{figure*}[ht!] \centering
\includegraphics[width=13cm]{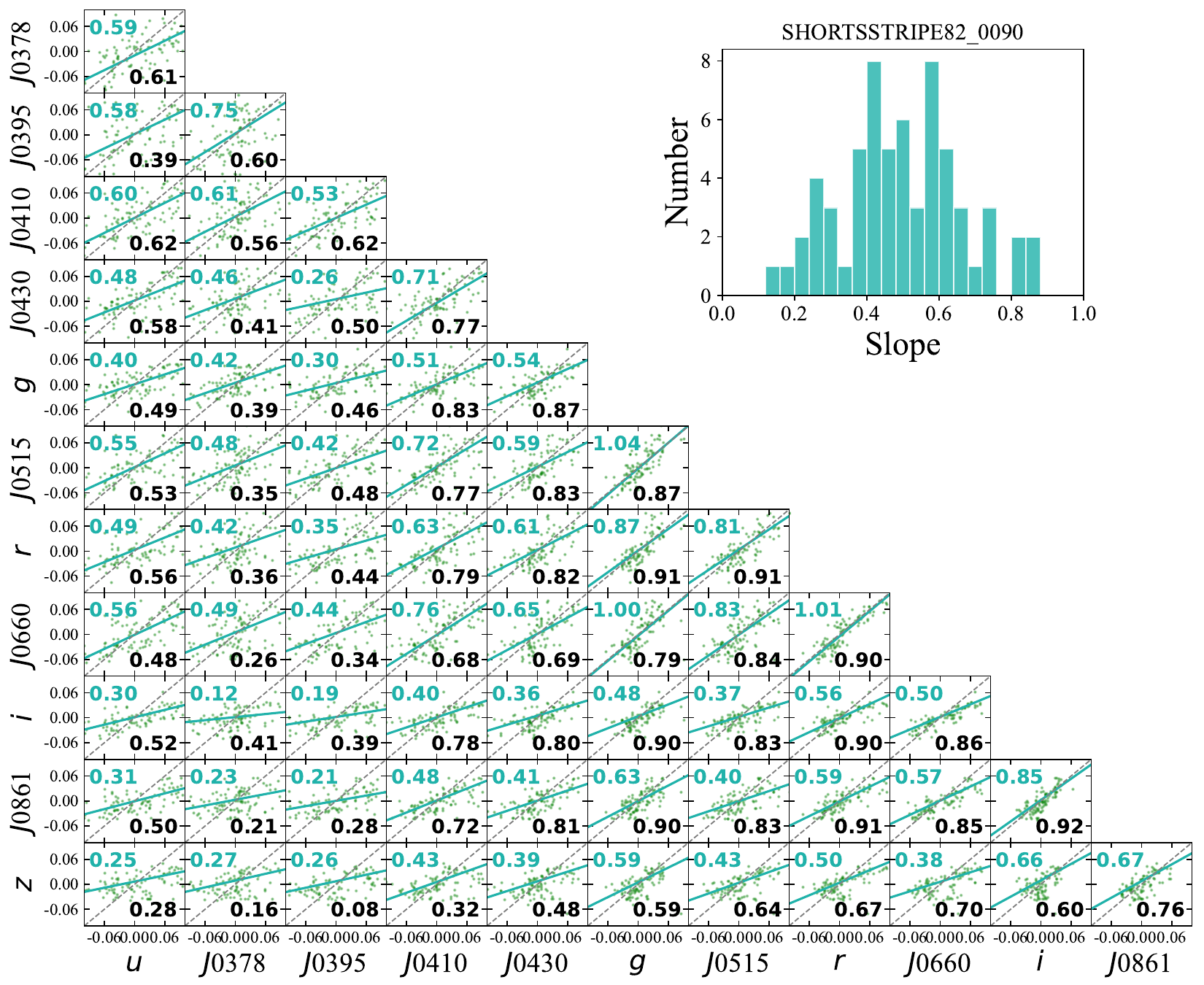}
\caption{{\small Same as Figure\,\ref{Fig:f6}, but for the correlation of the magnitude offsets, between the standards and USS photometry after minus zero-points for \texttt{tile\_ID=SHORTSSTRIPE82\_0090}, for each of the two bands. For each panel, the correlation coefficients are marked in the bottom-right corners. The linear fitting lines are shown as light-green lines, and the slopes of the line are marked in the top-left corners. The black-dashed lines denote $y=x$ in each panel. A histogram of the distribution of the slopes is shown in the top-right corner. }}
\label{Fig:f8}
\end{figure*}

\begin{figure*}[ht!]
   \centering
  \subfigure{\includegraphics[width=9cm]{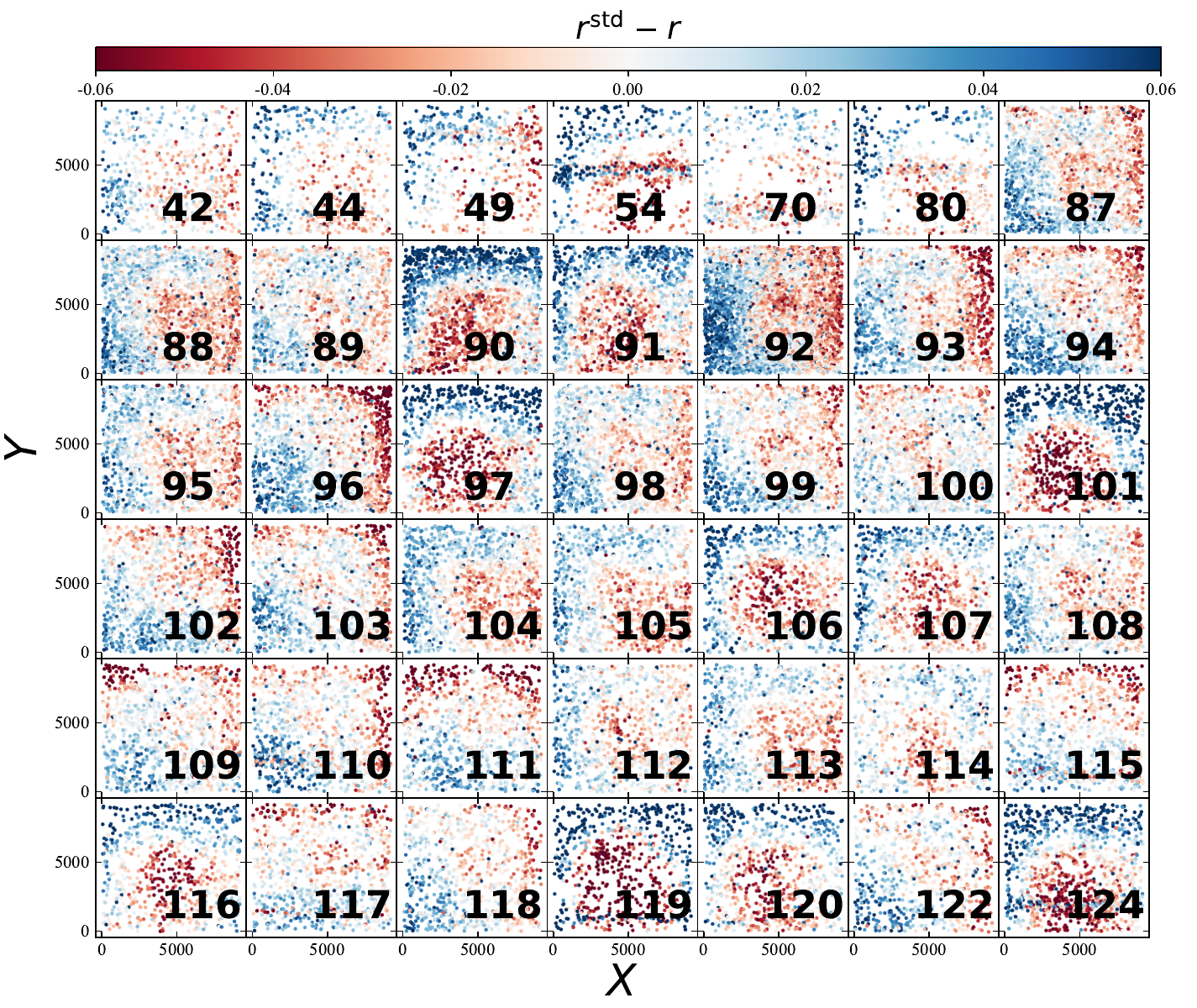}
            \includegraphics[width=9cm]{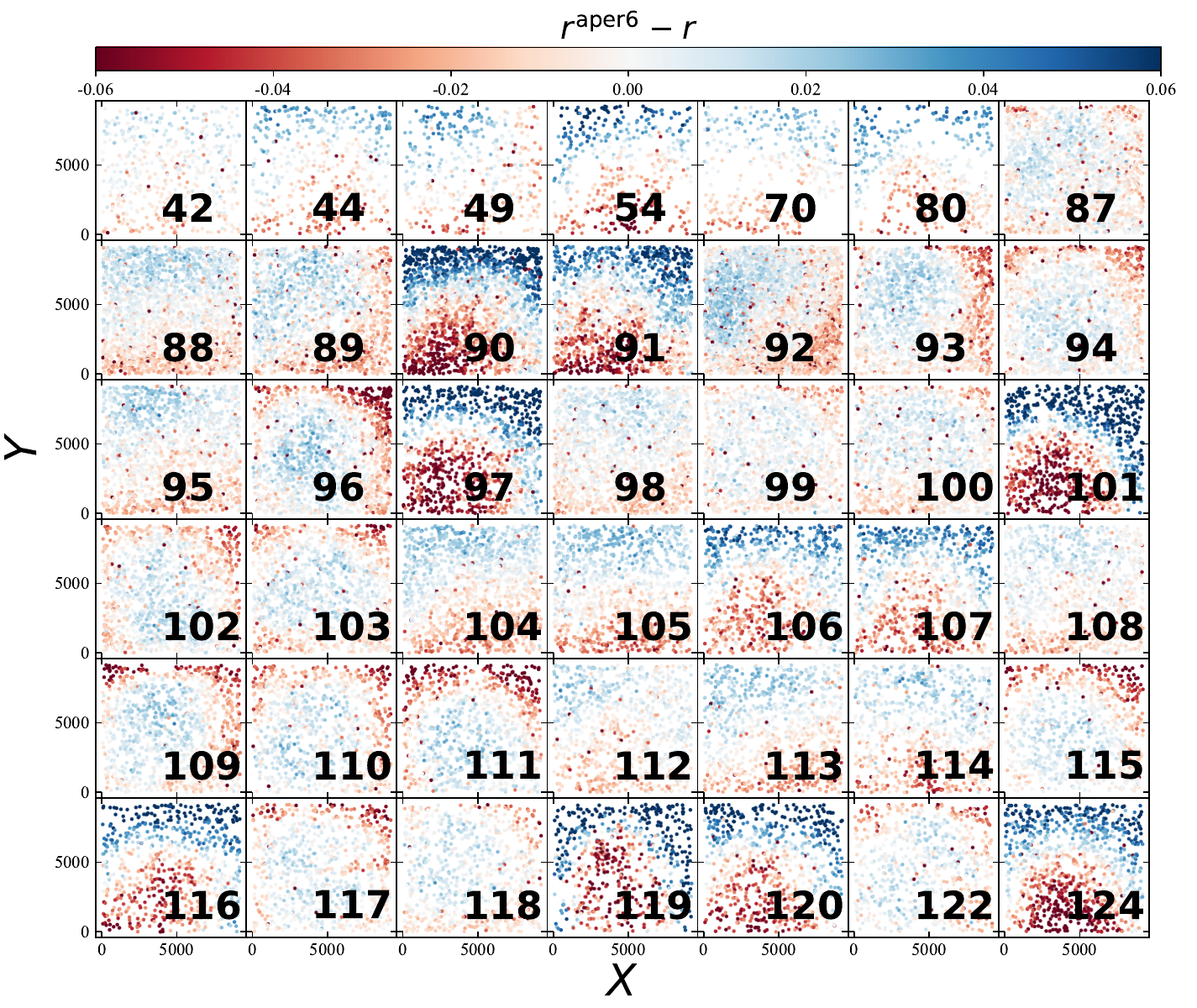}} \\
  \subfigure{\includegraphics[width=9cm]{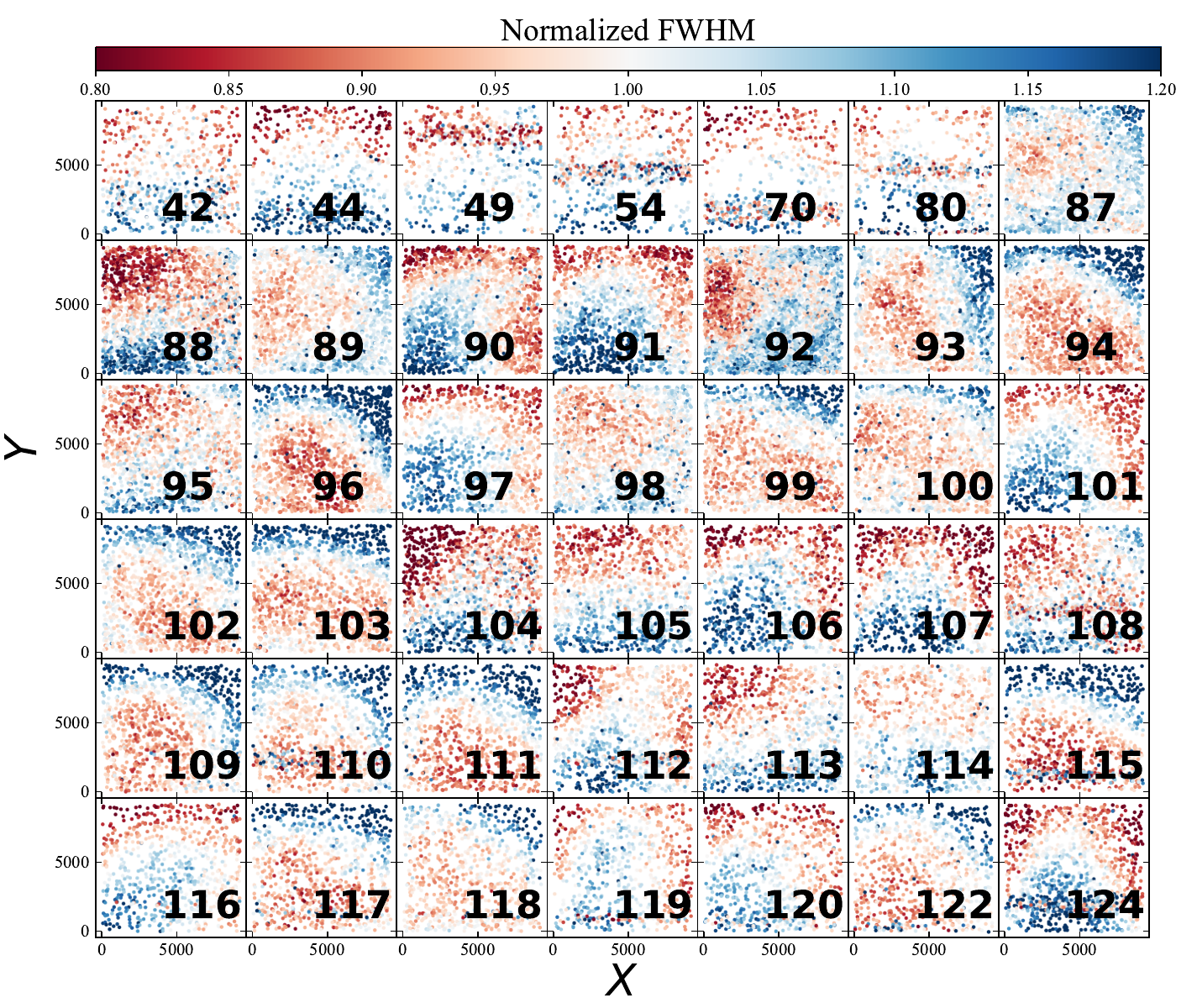}
            \includegraphics[width=9cm]{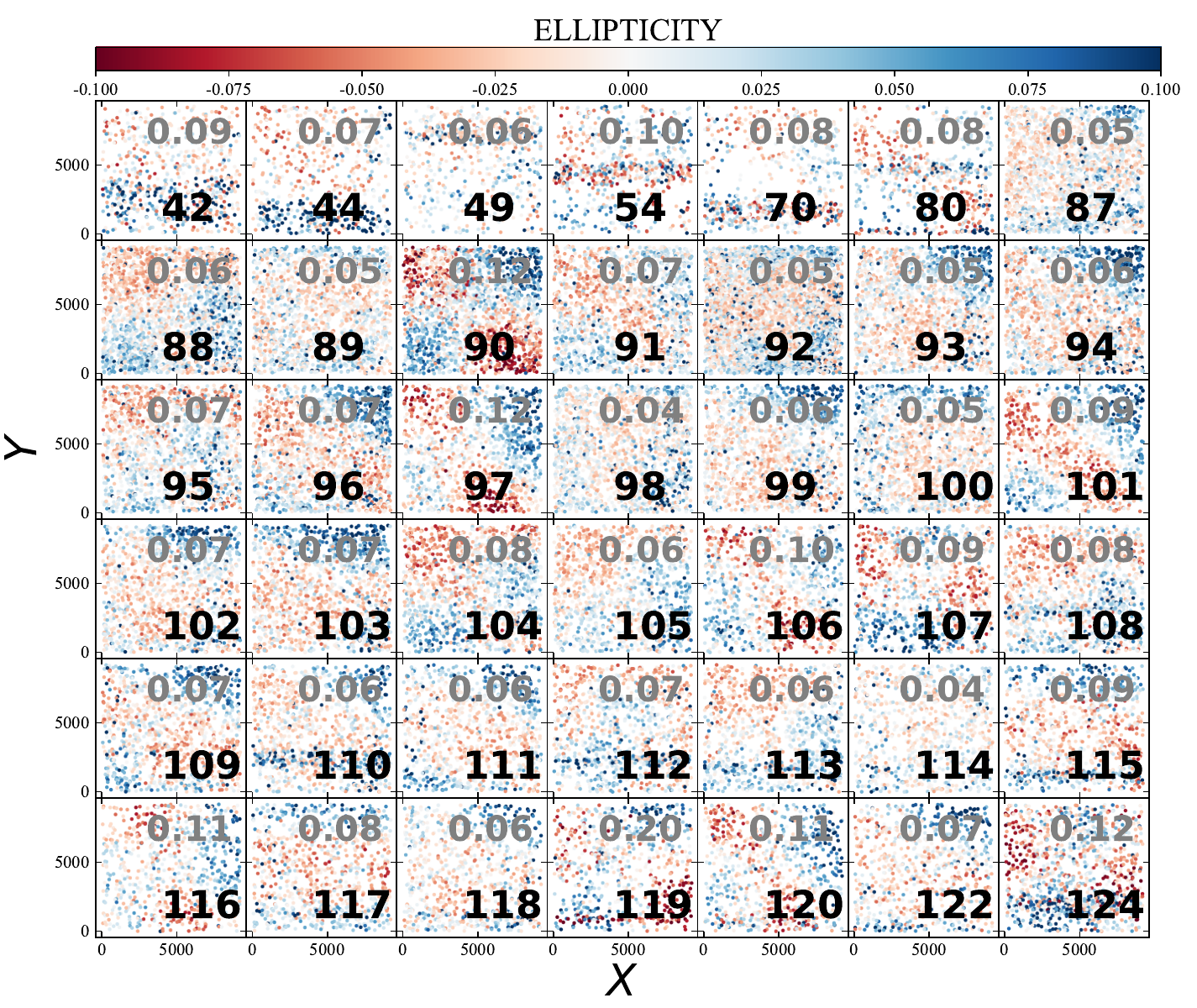}}
   \caption{{\small Examples of the spatial variations of the magnitude offsets between the XPSP standards and USS $r$-band photometry (the top-left panels), the magnitude offsets between the USS $r$-band magnitude with aperture 6 and aperture 3 after aperture correction (the top-right panels), the normalized FWHM (the bottom-left panels), and ellipticity (the bottom-right panels). For the bottom-right panels, the points are displayed after removing the median of all points in the tile, and the median value is marked (in gray) in the top-right corner for each panel. The \texttt{tile\_ID} is marked in each panel in black. The color bars are shown on the top, respectively.}}
  \label{Fig:f9}
\end{figure*}

\begin{figure*}[ht!] \centering
\includegraphics[width=14cm]{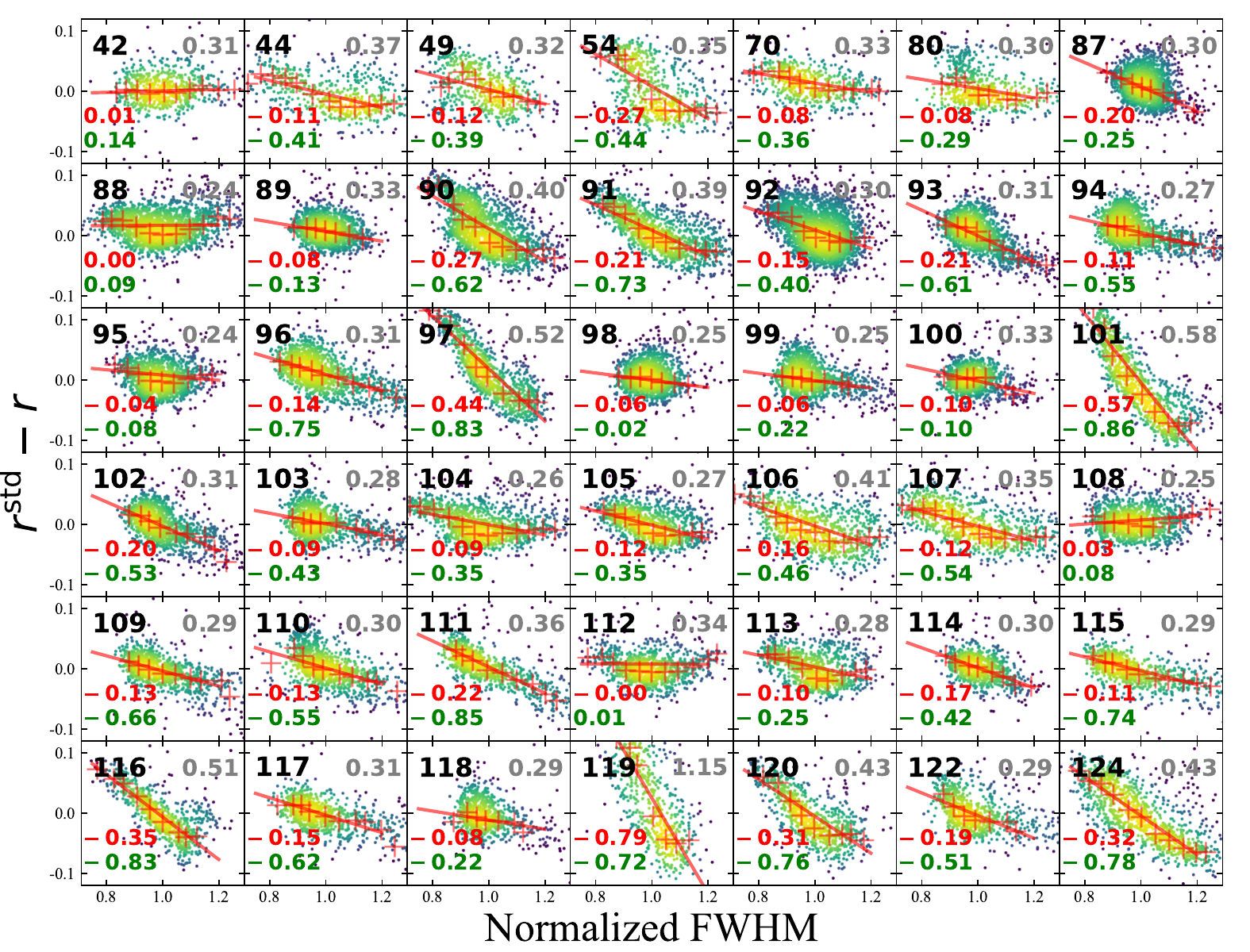}
\caption{{\small Examples of correlation plots of the magnitude offsets between the standards and the USS $r$-band photometry, as a function of the normalized FWHM. For each panel, the red plus signs denote the median values of the points, and the red lines denote the linear fitting result to the red pluses. The \texttt{tile\_ID} (black), the correlation coefficient (green), and the slope (red) are marked in each panel. The gray number in each panel represent the magnitude offset in the aperture-correction process.}}
\label{Fig:f10}
\end{figure*}

To determine the zero-point offsets of USS, we calculated the median value of the magnitude offsets between the BEST standards and the USS DR1 stars for each tile, based on the calibration stars (refer to the magnitude range in Figure\,\ref{Fig:f2}). We note that using only very bright sources or a large number of faint sources can introduce significant random errors to the zero-points. For example, the brighter the sources used, the fewer standard stars will be present in the tile, leading to a larger scatter in the zero-points. As the balance between high signal-to-noise ratio and a sufficient number of sources is achieved, the scatter in the zero-point will gradually stabilize and become more accurate (this work).

Figure\,\ref{Fig:f4} illustrates significant spatial variations in the zero-point offsets, especially for the blue filters, predominantly due to the higher uncertainties of the technique employed in the original USS DR1 calibration.

To quantitatively assess the systematic errors in the USS DR1 data, we generated histograms of the zero-point offsets between the BEST standards and USS, as shown in Figure\,\ref{Fig:f5}. By fitting these histograms to a Gaussian function, we calculated the scatter for each filter as shown in the first line in Table\,\ref{tab:pre}. These results align with \cite{2024arXiv240705004P}, and suggest an internal precision of about 7--8\,mmag for the $griz$ filters, 12--40\,mmag for the blue filters, and about 7--11\,mmag for the red filters within the USS DR1 photometry.

\begin{deluxetable*}{ccccccccccccc}[ht!]
\tablecaption{Internal Precision (in mmag) of the Photometric Calibration for the 12 USS Bands Before and 
After Re-calibration \label{tab:pre}}
\tablehead{
\colhead{Filters} & \colhead{$g$} & \colhead{$r$} & \colhead{$i$} & \colhead{$z$} & \colhead{$u$} & \colhead{$J0378$} & \colhead{$J0395$} & \colhead{$J0410$} & \colhead{$J0430$} & \colhead{$J0515$} & \colhead{$J0660$} & \colhead{$J0861$}}
\startdata
Before & $7.0$  & $8.1$  & $7.7$  & $7.1$  & $26.9$ & $28.1$ & $39.8$ & $11.9$ & $11.5$ & $7.0$  & $7.9$  & $6.5$ \\
After & $0.7$ & $0.7$ & $1.0$ & $1.0$ & $6.4$ & $4.6$ & $3.8$ & $2.0$ & $1.5$ & $1.1$ & $1.2$ & $0.6$ \\
\enddata
\end{deluxetable*}

In order to trace the systematic errors, the correlation between the zero-point offsets for each two-band filter pair are illustrated in Figure\,\ref{Fig:f6}. 
The systematics typically arise from at least two sources. For the blue-band pairs (involving the $u$, $J0378$, $J0395$, $J0410$, and $J0430$ filters), we estimated the slope through linear fitting. This slope varies between different band pairs. A common source of those systematic errors is the inadequate handling of stellar metallicity in the SL method. The slope value reflects the ratio of metallicity sensitivities between the two bands. For example, the slopes for the $J0378$- versus $J0395$-bands and the $J0430$- versus 
$J0410$-bands are close to unity, because the sensitivity to metallicity is nearly identical for the 
$J0378$- and $J0378$-bands, and the $J0430$- versus $J0410$-bands, respectively. Details on the metallicity sensitivity in the blue bands can be found in Figure\,A1 of \cite{2023ApJS..269...58X}. Conversely, the slope for the $J0430$- versus $u$- bands is 0.18, because the $u$-band is significantly more sensitive to metallicity than the $J0430$-band.
In addition, for the red-band filter pairs (e.g., $J0515$- versus $g$-bands), the zero-points typically align along the $y=x$ line, and present a high correlation. We believe that this systematic error is primarily due to systematic errors in the ATLAS Refcat2.

\subsection{Dependence of Magnitude Offsets on CCD Position} \label{sec:star}
The CCD position variations in the magnitude offsets between the XPSP standards and USS DR1 (after removing the zero-level constant for each tile across all 12 bands) are shown in Figure\,\ref{Fig:f7}. For each tile, there are significant CCD position-dependent systematic errors, reaching up to 50 milli-mag, with correlations observed across different bands. 

To quantitatively analyze this correlation, taking the $r$-band as an example, we randomly selected one-tile (\texttt{SHORTS-STRIPE82\_0090}) from all tiles where the number of XPSP standards exceeds 1000. The correlation of magnitude offsets between each pair of bands for \texttt{SHORTS-STRIPE82\_0090} is plotted in Figure\,\ref{Fig:f8}. Within the same field, magnitude offsets between different bands present correlations, with 80\% of the band pairs having a linear correlation coefficient greater than 0.4. For any two bands, linear fitting is performed, and it reveals that 74\% of the tiles have a slope exceeding 0.4. For example, in the $\Delta J660$ versus $\Delta r$ panel, points are distributed along the $y=x$ line, with a correlation coefficient of 0.90 and a slope of 1.01. We then plotted the FWHM correlation for each pair of bands, as shown in Figure\,\ref{Fig:A3}. For each band pair, all the points are distributed along the $y=x$ line. Furthermore, we found strong correlations in the FWHM across different bands, and the magnitude offsets and FWHM for each band pair. We believe that the systematic errors primarily originate from the aperture-correction process. Two other tiles, \texttt{SHORTS-STRIPE82\_0096} and \texttt{SHORTS-STRIPE82\_0100}, exhibit the same phenomenon.

\begin{figure*}[ht!] \centering
  \subfigure{\includegraphics[width=8.8cm]{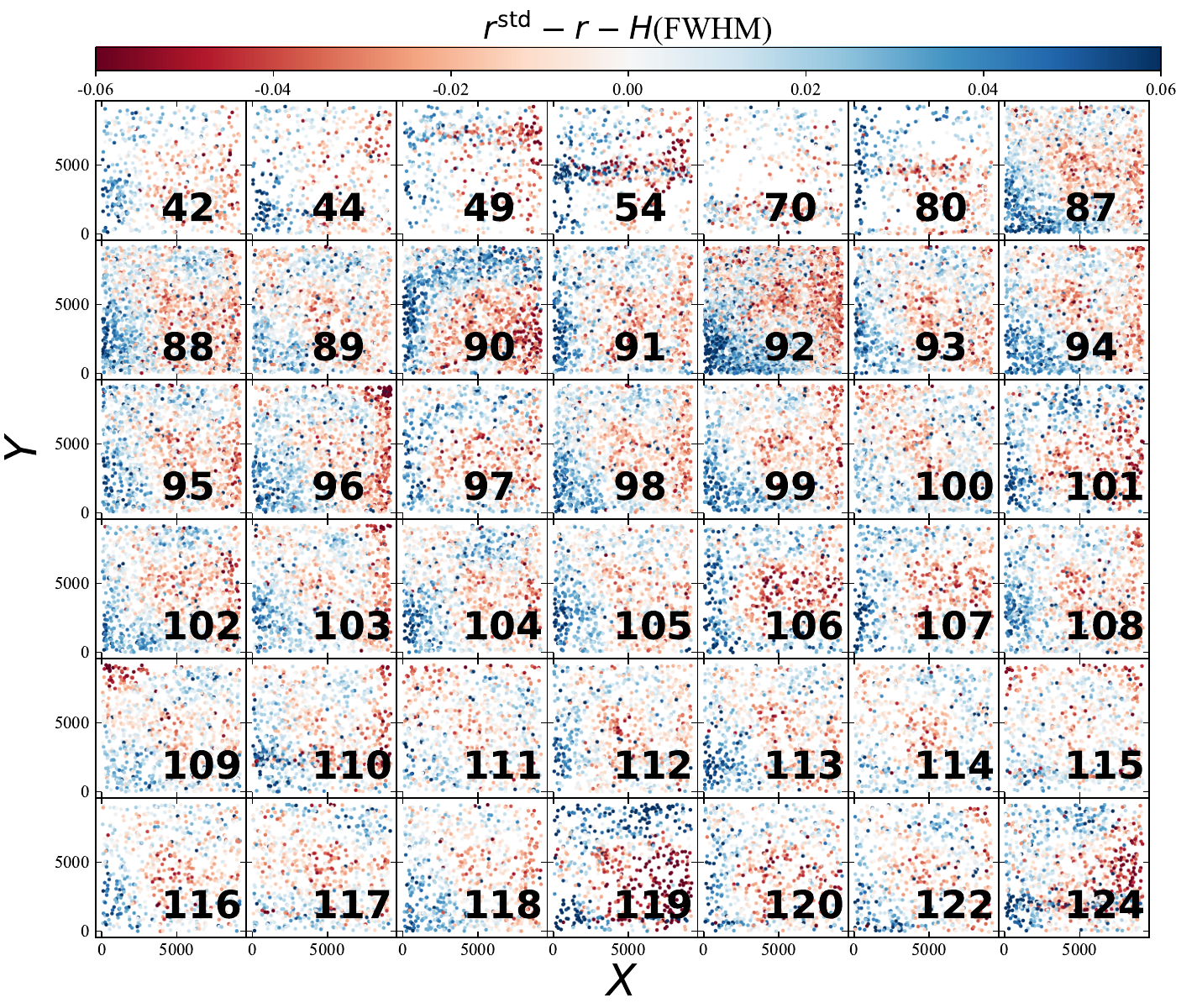}
             \includegraphics[width=8.8cm]{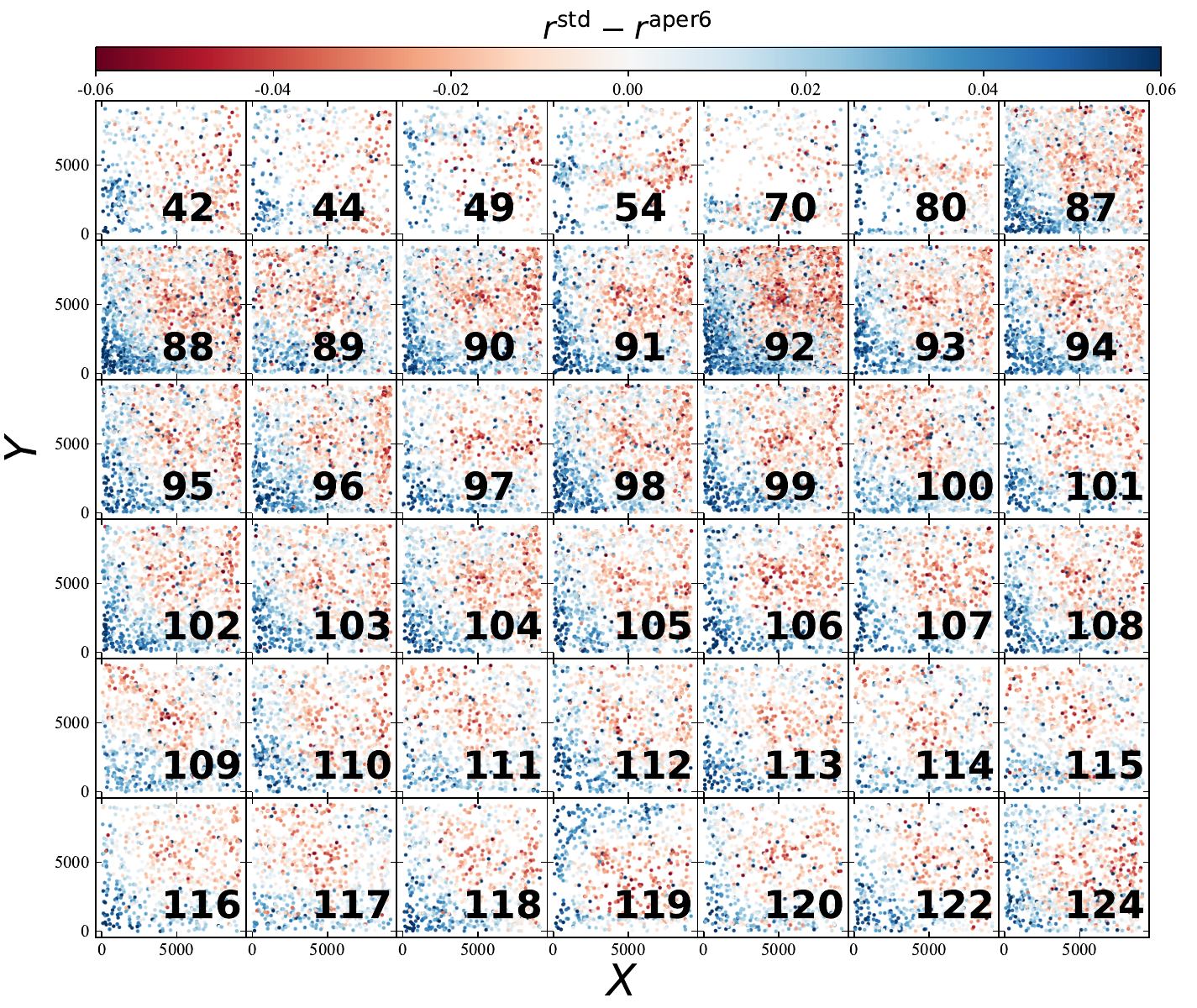}}
\caption{{\small Examples of the stellar flat-field in the $r$-band. Left panels: Spatial variations of the magnitude offsets between the standards and USS $r$-band photometry after FWHM-dependent systematic errors correction using a second-order two-dimensional polynomial (with six free parameters) fitting. The \texttt{tile\_ID} is marked in each panel in black. Right panels: Spatial variations of the magnitude offsets between the standards and USS $r$-band magnitude with \texttt{APER\_6}. The color bars are shown on the top.}}
\label{Fig:f11}
\end{figure*}

\begin{figure*}[ht!] \centering
  \subfigure{\includegraphics[width=9cm]{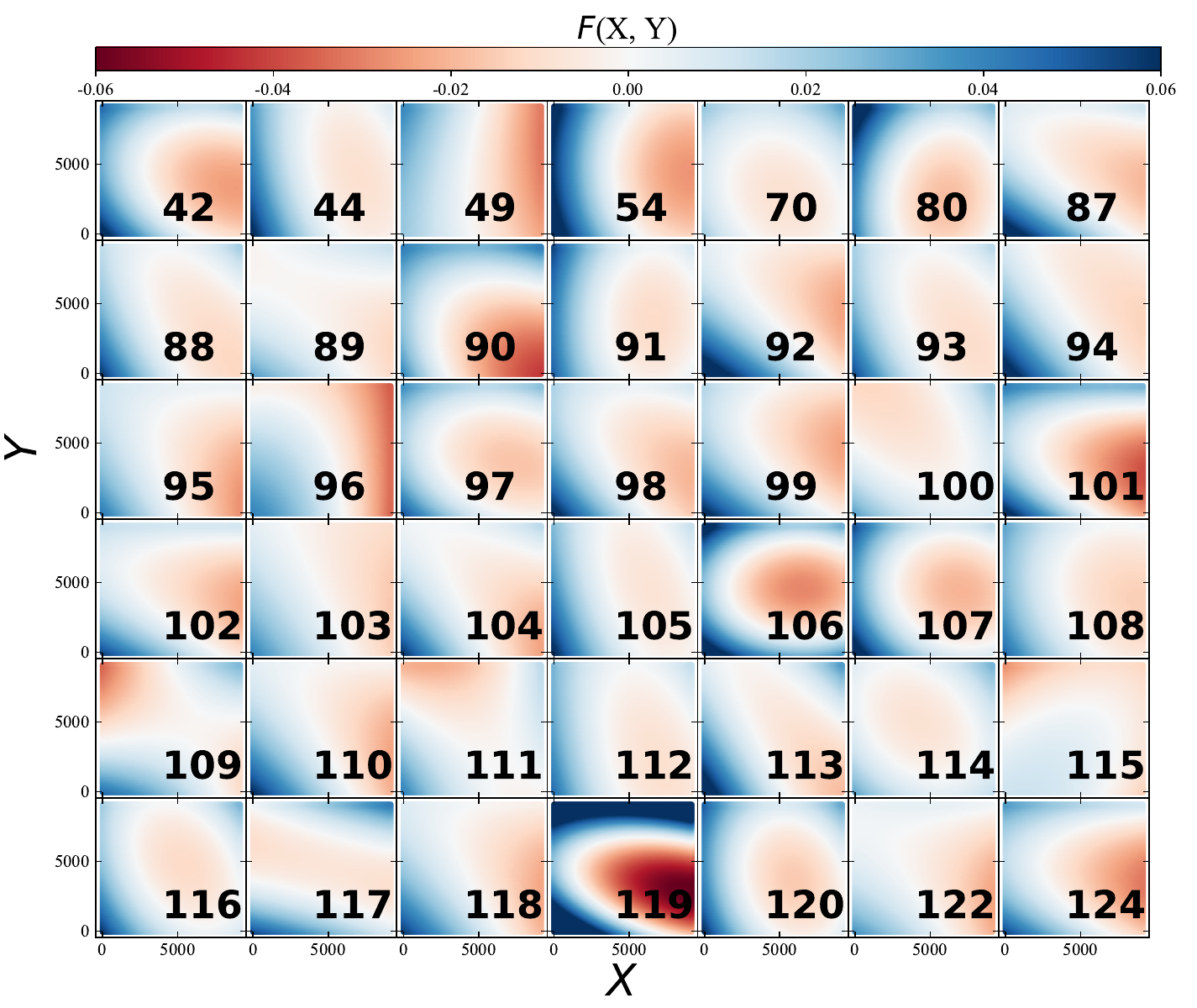}
  \includegraphics[width=9cm]{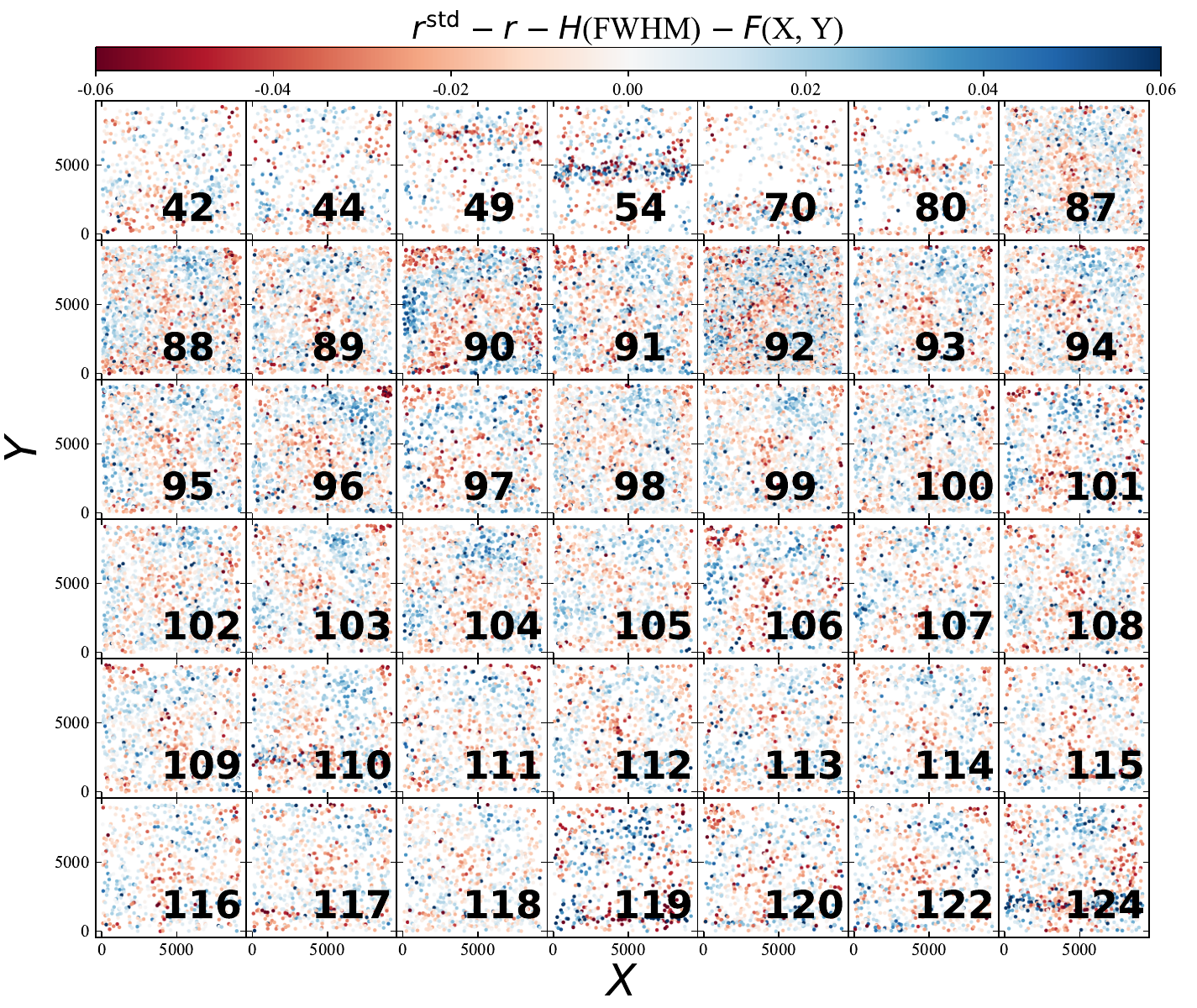}} \\
\caption{{\small Examples of the medium-scale stellar flat-field in the $r$-band. The color bars are shown on the top. Left panels: Spatial variations of the magnitude offsets between the standard and USS photometry after FWHM- and large-scale stellar flat-field correction. The \texttt{tile\_ID} is marked in each panel in black. Right panels: Residuals after application of the second-order two-dimensional polynomial.}}
\label{Fig:f12}
\end{figure*}

\begin{figure*}[ht!] \centering
\subfigure{\includegraphics[width=9.8cm]{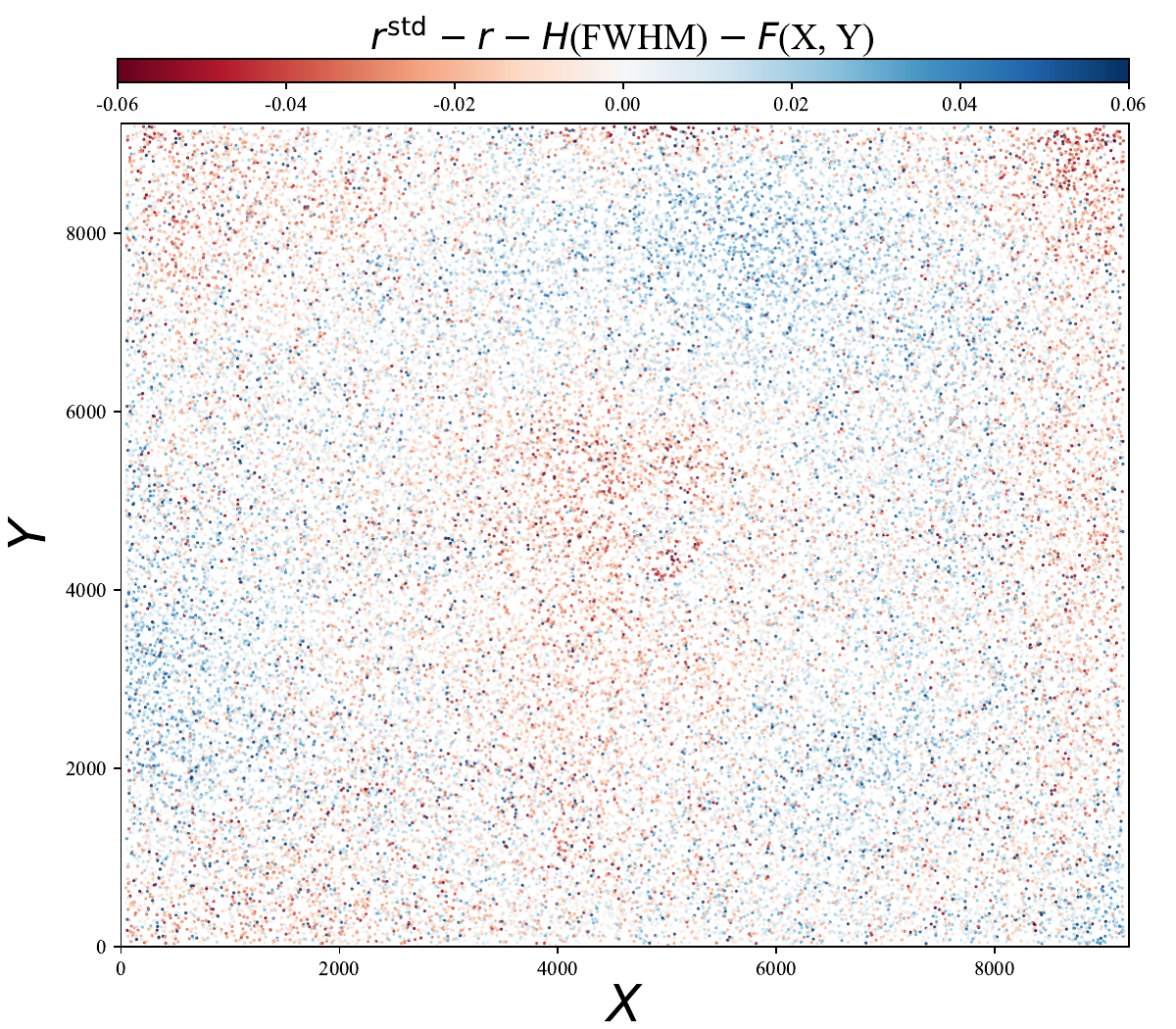}
           \includegraphics[width=8.8cm]{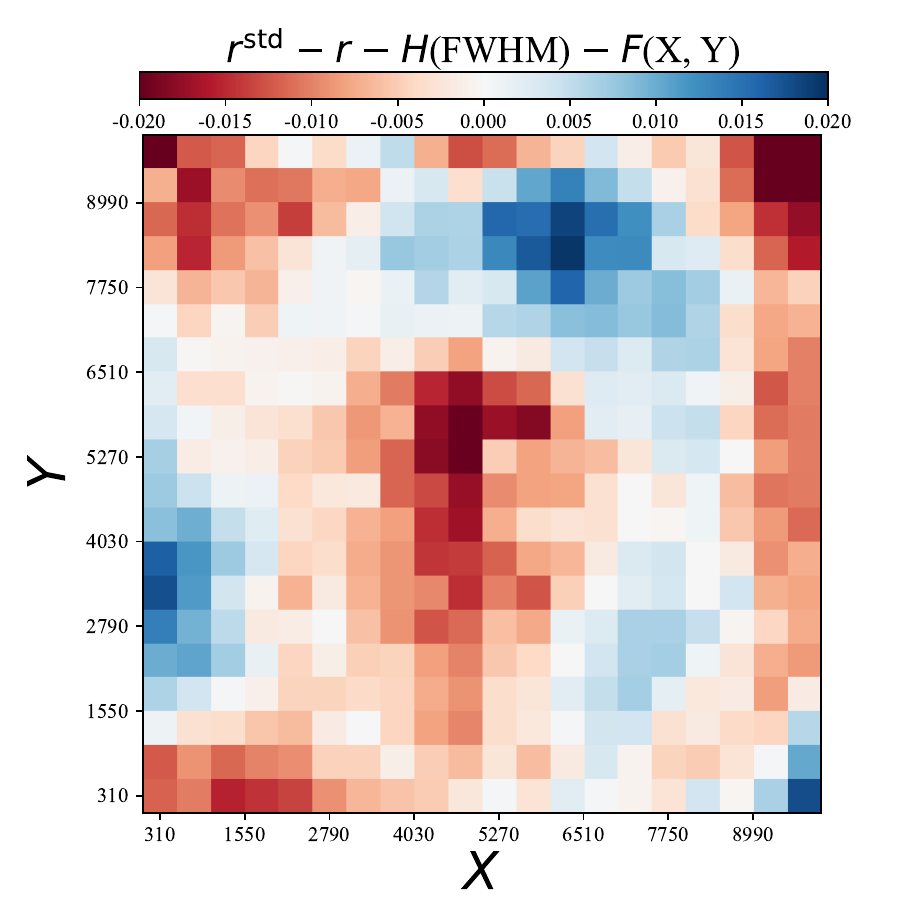}} \\
\caption{{\small Spatial variations of the second-order two-dimensional polynomial fitting residuals of 42 tiles (left panel) and the results after $20\times 20$ binning (right panel). The color bars are shown on the top.}}
\label{Fig:A1}
\end{figure*}

\begin{figure*}[ht!] \centering
\resizebox{\hsize}{!}{\includegraphics{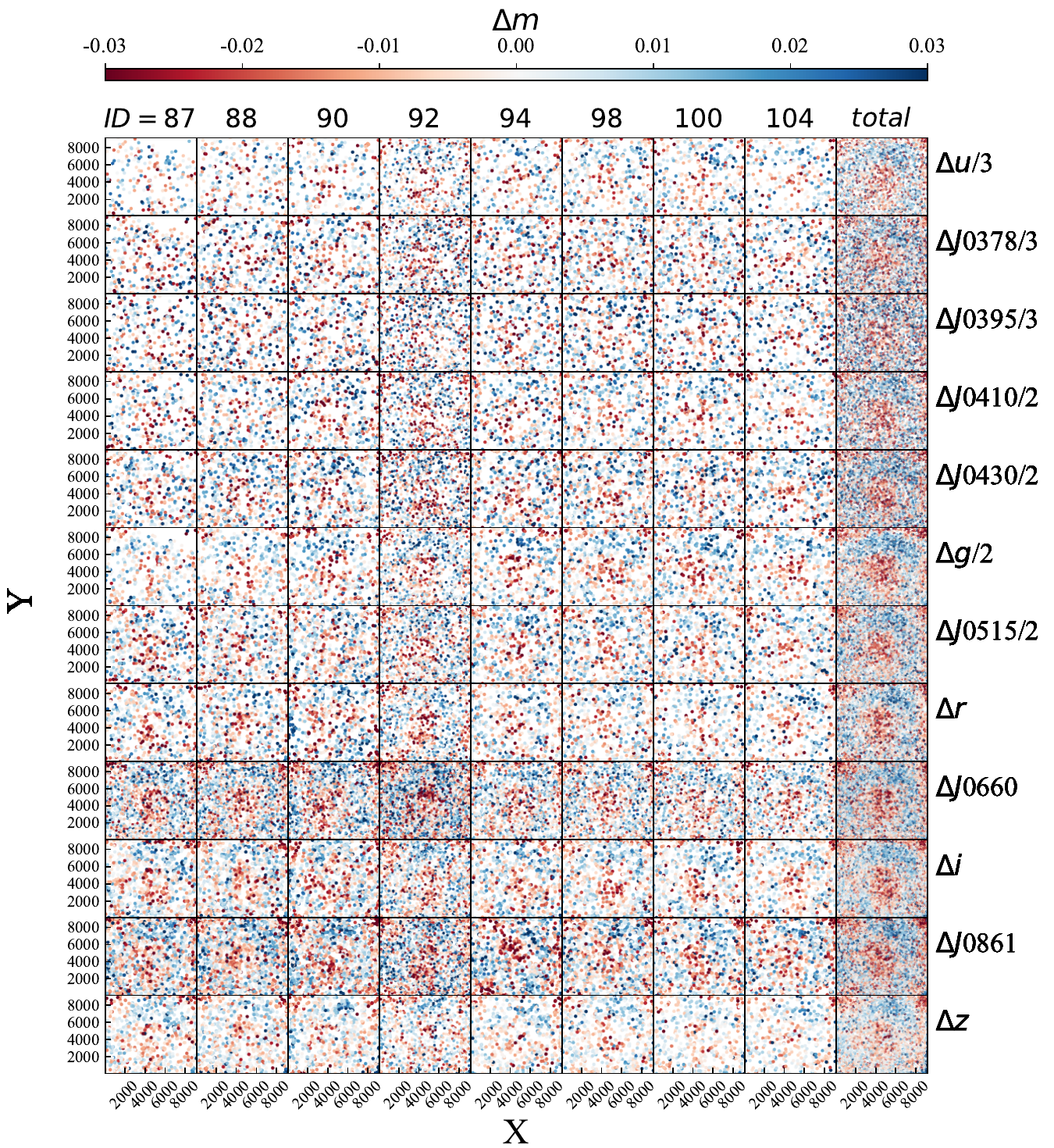}}
\caption{{\small An example showing the CCD spatial distribution of residuals after applying the second-order two-dimensional polynomial correction. The panels, arranged from top to bottom, display the panels for the $u$-, 
$J0378$-, $J0395$-, $J0410$-, $J0430$-, $g$-, $J0515$-, $r$-, $J0660$-, $i$-, $J0861$-, and $z$-bands, respectively. The bands are labeled at the right, and a color bar is provided on the top. From left to right, the first eight columns represent the selected eight tiles (\texttt{SHORTS-STRIPE82\_0087}, \texttt{SHORTS-STRIPE82\_0088}, \texttt{SHORTS-STRIPE82\_0090}, \texttt{SHORTS-STRIPE82\_0092}, \texttt{SHORTS-STRIPE82\_0094}, \texttt{SHORTS-STRIPE82\_0098}, \texttt{SHORTS-STRIPE82\_0100}, \texttt{SHORTS-STRIPE82\_0104}), with the final column showing the combined residuals of these eight tiles. The tile number is marked between the first-row panels and the color bar. To clearly show the spatial structure, the results for the $u$-, $J0378$-, $J0395$-, $J0410$-, $J0430$-, $g$-, and $J0515$-bands are compressed two to three times. }}
\label{Fig:f20}
\end{figure*}

\begin{figure*}[ht!] \centering
\resizebox{\hsize}{!}{\includegraphics{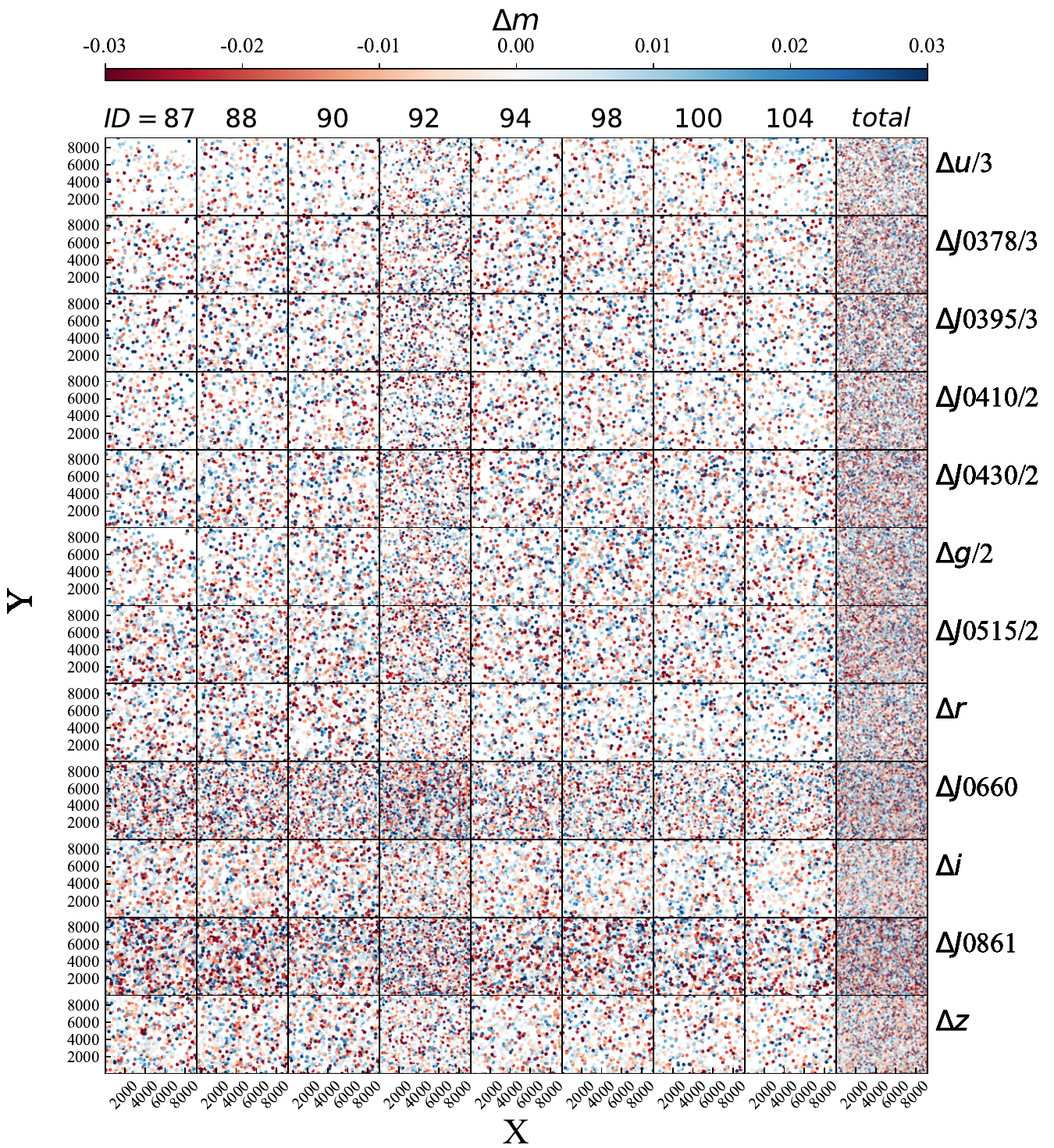}}
\caption{{\small Similar to Figure\,\ref{Fig:f20}, but showing the results after applying both the second-order two-dimensional polynomial and numerical stellar flat-field corrections.}}
\label{Fig:f13}
\end{figure*}

\begin{figure*}[ht!] \centering
\resizebox{\hsize}{!}{\includegraphics{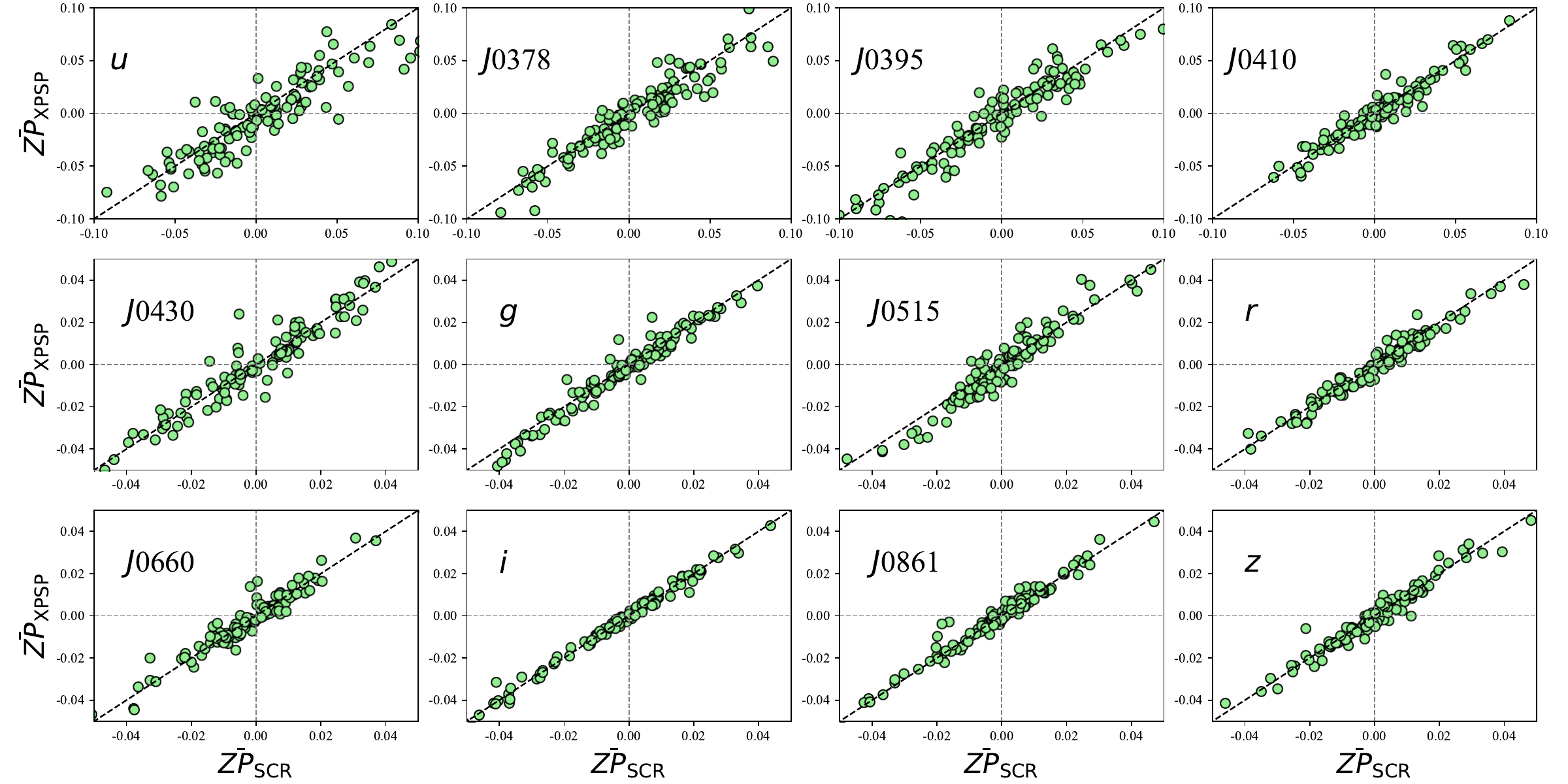}}
\caption{{\small Correlation plots between the zero-point offsets predicted by the XPSP and the SCR standard stars for all bands. For each panel, the bands are marked in the top-left corner. The black-dashed lines denote $y=x$ in each panel; the gray-dashed lines denote the zero level.}}
\label{Fig:f14}
\end{figure*}

We next selected 42 tiles with more than 700 standard stars from all $r$-band observations, and plotted the distribution of the difference between the XPSP and the USS magnitudes, the difference between \texttt{APER\_6} and USS magnitudes, normalized FWHM, and ellipticity across the CCD space, as shown in Figure\,\ref{Fig:f9}. We found that these CCD position-dependent systematic errors strongly correlate with the distribution of the difference between the \texttt{APER\_6} and USS magnitudes in CCD space, show a strong negative correlation with the spatial distribution of FWHM, and exhibit almost no correlation with ellipticity  or the astrometric offsets with Gaia (please refer to Figure\,\ref{Fig:A2}). This indicates that the aperture size used for USS magnitudes in the photometric process is too small to fully capture the total flux of sources having large FWHM. Although CCD position-dependent systematic errors caused by selecting apertures that are too large or too small should be well-corrected during the aperture-correction process, this appears to not be the case. We remind the reader that the USS magnitudes here refer to \texttt{APER\_3} magnitudes after aperture correction, as mentioned in Section\,\ref{sec:uss}.

To further quantify this correlation, we plotted the distribution of the difference between the XPSP and USS magnitudes, as a function of normalized FWHM, across different bands for tile \texttt{SHORTS-STRIPE82\_0090}, as shown in Figure\,\ref{Fig:f10}. The linear correlation coefficient is as high as 0.86, and the linear fitting results, $H(\rm FWHM)$, with a slope of up to 0.79, are also noted in the figure. Since the absolute reference value of the FWHM is unavailable, we used the normalized FWHM for analysis here. As is well known, the absolute FWHM for a tile is positively correlated with the median value of the aperture-correction value, In other words, a tile with larger FWHM will often have larger aperture corrections. Although the absolute FWHM is not entirely equivalent to the median value of the aperture correction for a given tile, the correlation between them is still valuable for analysis. As shown in Figure\,\ref{Fig:f10}, panels with larger aperture corrections exhibit more pronounced variations in stellar magnitude differences with changes in the normalized FWHM (e.g., panels 119 and 101). In contrast, the variations are more subdued in plots with larger aperture corrections (e.g., panels 99 and 118).

Using the linear fitting results, we then corrected the FWHM-dependent systematic errors for all 42 tiles. The corrected results are shown in the left panel of Figure\,\ref{Fig:f11}. After correcting, the residuals still exhibit a moderate dependence on CCD position. The spatial distribution of the residual errors closely mirrors that of the difference between the XPSP and USS \texttt{APER\_6} magnitudes, shown in the right panel of Figure\,\ref{Fig:f11}. The residual errors are primarily attributed to a different flat-field correction.

Furthermore, for each tile, we fitted the residuals after correcting for FWHM-dependent systematic errors using a second-order two-dimensional polynomial (with six free parameters), as a function of $X$ and $Y$. The results of second-order two-dimensional polynomial fitting, $F(X, Y)$, for all 42 tiles is shown in the left panel of Figure\,\ref{Fig:f12}. We then applied these polynomials to each tile to correct the residual errors; the corrected residuals are presented in the right panel of Figure\,\ref{Fig:f12}. It is notable that, for each panel, the spatial structure of the residuals exhibits an intermediate-scale pattern between large-scale and small-scale flat fields, and remains nearly consistent across different tiles. To illustrate this intermediate-scale structure more clearly, we combined the residuals of the 42 tiles, and plotted the results in the left panel of Figure\,\ref{Fig:A1}. To avoid overcrowding, 400 bins are uniformly selected in the CCD space, and the median value of the residuals within each bin ais used as the value for that bin. This smoothed result is displayed in the right panel of Figure\,\ref{Fig:A1}. A mottling pattern, reaching up to 40--50\,mmag for the $u$-, $J0378$-, $J0395$-, $J0410$- and $J0430$-band filters, and 10--20\,mmag for other filters, is clearly visible. This intermediate-scale structure is widely found in modern survey data, such as SAGES DR1 \citep[refer to Figure\,7 in][]{xiao}, J-PLUS DR3 \citep[depicted in Figure\,15 in][]{2023ApJS..269...58X}, and S-PLUS iDR4 \citep[see Figure\,10 in][]{2024ApJS..271...41X}. A detailed 
description of this intermediate-scale structure is presented in Section\,\ref{mottling} below.

\begin{figure*}[ht!] \centering
\resizebox{\hsize}{!}{\includegraphics{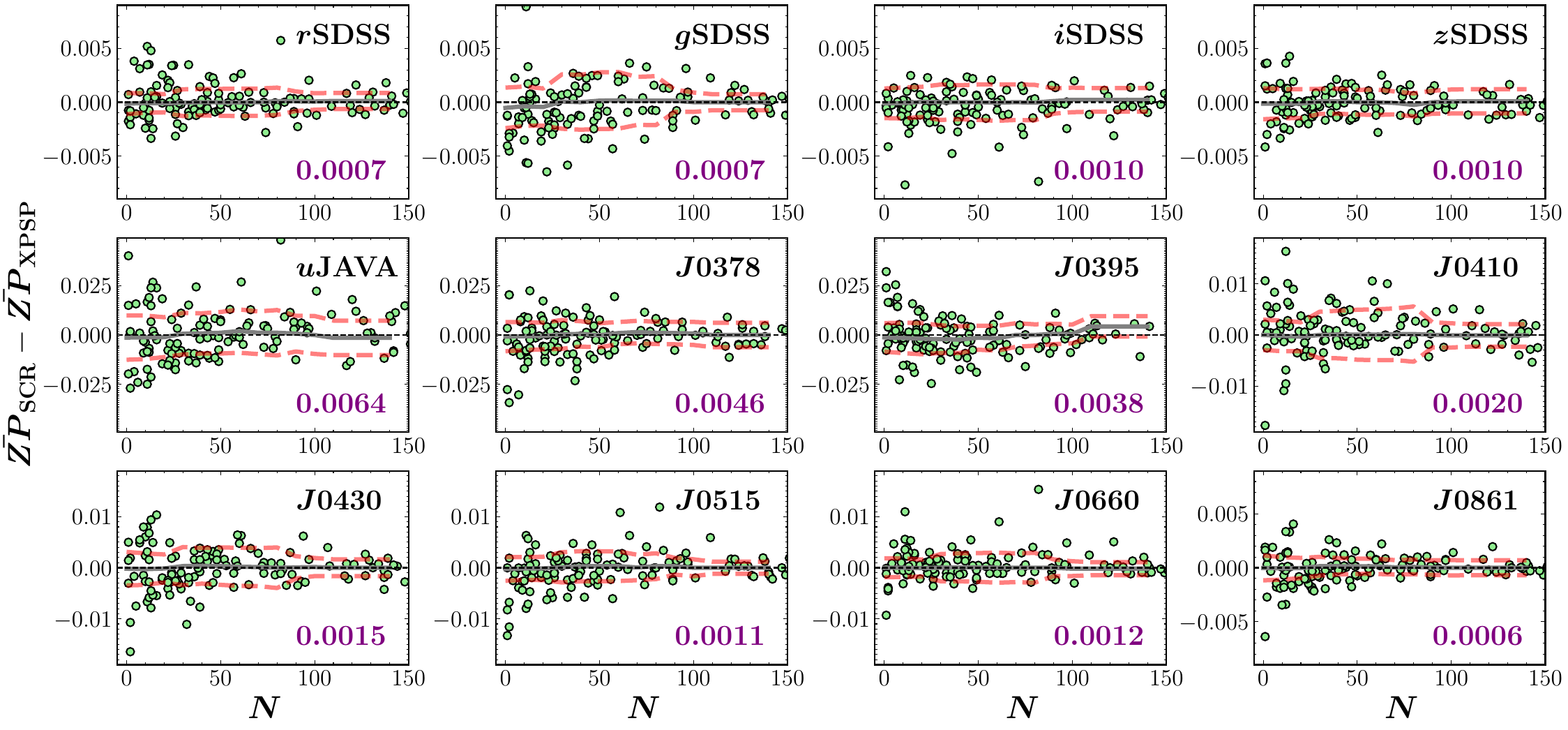}}
\caption{{\small Comparison of the zero-points for each of the SCR and XPSP standard stars for all 12 bands, as a function of star numbers (denoted as $N$) in each tile. The bands are marked in each panel. The green points show the difference of the zero-points. Their median values and standard deviations are estimated using Gaussian fitting with a running width of 15 stars and a running step of 1 star, and are indicated by the gray-solid and red-dashed curves, respectively.
The median value of the standard deviations, for $N$ ranging from 100 to 150, is labeled in each panel. The black-dashed lines represent the zero-residual line.
The distribution of zero-point differences is not strictly symmetric about zero at intermediate and/or large $N$, primarily due to the limited number of fields in common. However, the decreasing trend of scatter with increasing $N$ is clearly seen in all bands.}}
\label{Fig:f15}
\end{figure*}

\section{Systematic Error Corrections} \label{sec:correct}
Section\,\ref{sec:vali} highlights the spatially dependent systematic errors in the zero-points of the USS photometry, as well as systematics resulting mainly from different aperture and flat-field corrections. Both the systematic errors from the aperture and the flat-field correction process can be modeled as functions of CCD position ($X$, $Y$).

To accurately correct for these systematic errors in the USS data, we first applied a second-order two-dimensional polynomial, as a function of CCD position ($X$ and $Y$), fit to the difference between the standard star magnitudes ($m^{\rm std}$), including the SCR and the XPSP standards, and the USS magnitudes ($m$) for each tile. For the $i$th-band and $j$th tile, the second-order two-dimensional polynomial is expressed as:
\begin{equation}
\nonumber
{{\mathcal F}_{i, j}} = a_{i, j}^{0}\cdot X^2 + a_{i, j}^{1}\cdot Y^2 + a_{i, j}^{2}\cdot X\cdot Y + a_{i, j}^{3}\cdot X + a_{i, j}^{4}\cdot Y + a_{i, j}^{5}~.
\end{equation}
Here, $a_{i, j}^{0}$, $a_{i, j}^{1}$, $a_{i, j}^{2}$, $a_{i, j}^{3}$, $a_{i, j}^{4}$, and $a_{i, j}^{5}$ represent six free parameters for the $i$th band and $j$th tile.

These polynomial coefficients were then used to correct for the large-scale structures in CCD spatial and zero-point offsets in each tile. We required that, for each band and tile, the number of standard stars exceeded 20, ensuring more data points than the number of polynomial coefficients.

After applying the second-order two-dimensional polynomial correction, Figure\,\ref{Fig:f20} shows the CCD spatial distribution of residuals for eight tiles, each with a relatively high number of calibration stars, across all 12 bands. Different tiles exhibit a consistent mottling structure within the same band, and different bands display a similar pattern for the same tile. This suggests that the medium-scale flat-field structure remains consistent across bands and over short time scales, and also indicates that the second-order polynomial correction effectively corrects CCD position-dependent systematic errors from both different aperture and flat-field corrections in the USS DR1 photometric data.

Although a carefully modeled radial function, constructed with a basis function of $f({\sqrt {X^2 + Y^2}})$, could correct part of the mottling pattern in the USS DR1 data, a more effective approach for general cases, however, would be to use the numerical stellar flat-field correction method discussed in \cite{2024ApJS..271...41X}. The medium-scale CCD position-dependent systematic errors are then corrected using the numerical stellar flat-field correction method for each tile.

The difference between the XPSP standards and the USS magnitudes, after applying both second-order two-dimensional polynomial and numerical stellar flat-field corrections, is shown in Figure\,\ref{Fig:f13}. It is evident that, following re-calibration, the photometric data are significantly more uniform. For example, for \texttt{SHORTS-STRIPE82\_0092}, the consistency of the 
USS $r$-band photometry improved from 27\,mmag to 8.9\,mmag.

 When using the re-calibrated USS DR1 photometric data, one can download the FITS file \texttt{Recalibrated\_USS\_DR1\_Photometry.fits} from \href{https://nadc.china-vo.org/res/r101504/}{doi: 10.12149/101503}\footnote{\url{https://nadc.china-vo.org/res/r101504/}}. This file contains 96 columns, with each band containing the following 8 columns: right ascension (\texttt{RA\_band}), declination (\texttt{DEC\_band}), original aperture magnitude after aperture correction (\texttt{band\_PStotal}), magnitude error (\texttt{band\_err\_PStotal}), photometric quality flag (\texttt{band\_flag}), stellar probability (\texttt{band\_class\_star}), aperture magnitude after second-order two-dimensional polynomial stellar flat-field correction (\texttt{band\_PStotal\_D2O2}), and the final calibrated aperture magnitude (\texttt{band\_PStotal\_corr\_final}) that accounts for numerical stellar flat-field correction.

\section{Re-calibration Precision in the Zero-points} \label{sec:s5}
\begin{figure}[ht!] \centering
\resizebox{\hsize}{!}{\includegraphics{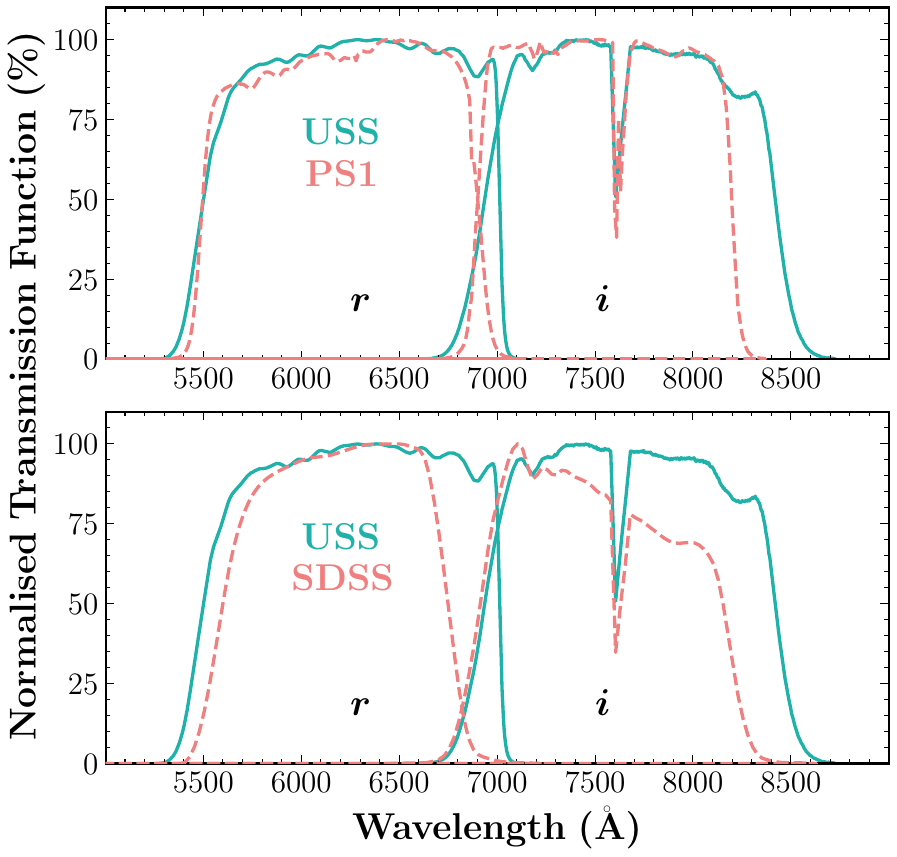}}
\caption{{\small Normalized transmission functions, as a function of wavelength, for the $r$- and $i$-bands. The labels are marked in each panel. The green curves denote the USS system transmission curves. For the top panel, the pink-dashed curves denote the PS1 photometric transmission curves. For the bottom panel, the pink-dashed curves denote the SDSS system transmission curves.}}
\label{Fig:f16}
\end{figure}

\begin{figure}[ht!] \centering
\resizebox{\hsize}{!}{\includegraphics{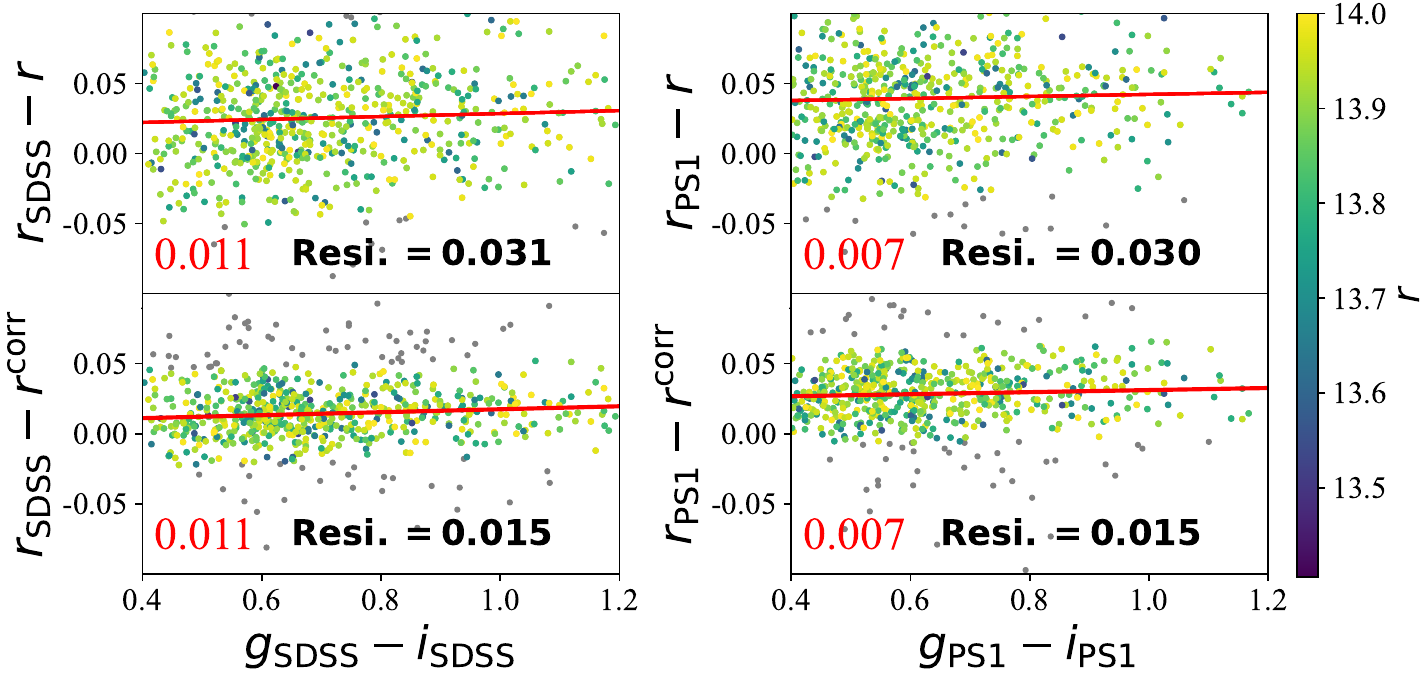}}
\caption{{\small An example showing the difference between the SDSS and USS $r$-band magnitudes before (the top-left panel) and after re-calibration (the bottom-left panel), and the PS1 and USS $r$-band magnitudes before (the top-right panel) and after
re-calibration (the bottom-right panel). For each panel, the red line represents the result of linear fitting;the slope (red) and the standard deviation of residuals (black) are labeled. A color bar coding the USS $r$-band magnitudes is shown at the right.}}
\label{Fig:f17}
\end{figure}

\begin{figure*}[ht!] \centering
\resizebox{\hsize}{!}{\includegraphics{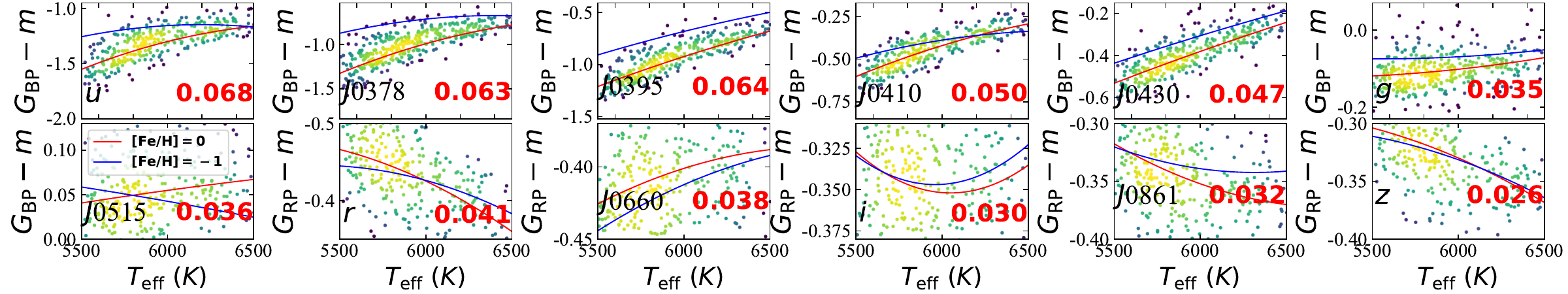}}
\resizebox{\hsize}{!}{\includegraphics{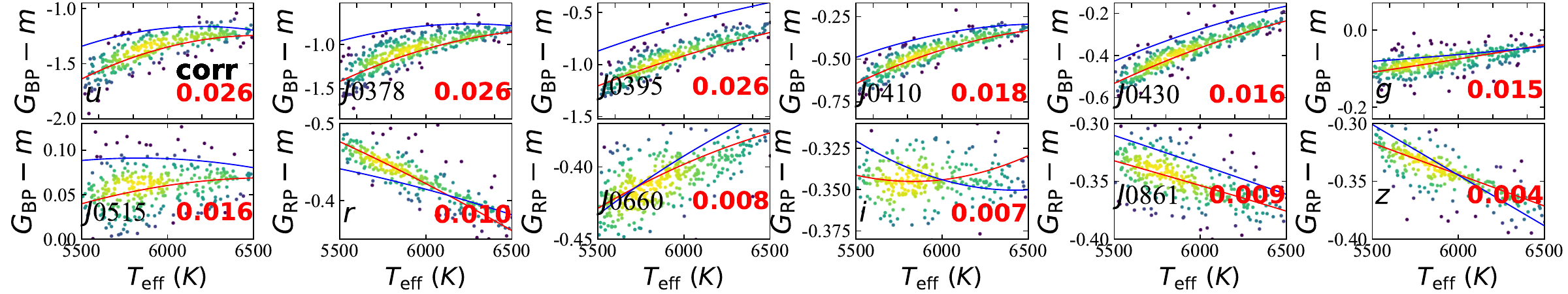}}
\resizebox{\hsize}{!}{\includegraphics{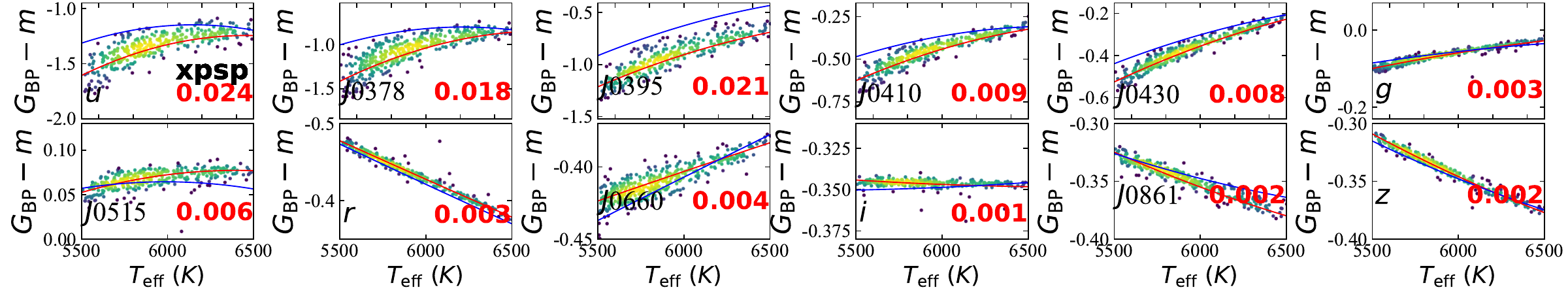}}
\caption{{\small Polynomial fits of the intrinsic colors, with respect to $T_{\rm eff}$ and $\rm [Fe/H] $, for the selected stars.  For each panel, the colors of the points represent the logarithm of the number density, calculated using Gaussian kernel density estimation.} The fit results after 3-$\sigma$ clipping are shown in each panel, with the red and blue curves representing results for $\rm [Fe/H] = 0 $ and $\rm [Fe/H] = -1$, respectively. The fitting residuals are labeled in red. The stellar color predicted by $G_{\rm BP/RP}-m$, where $m$ denotes the USS magnitude before re-calibration (the top 12 panels), the USS magnitude after re-calibration (the middle 12 panels), and the USS XPSP magnitude (the bottom 12 panels).}
\label{Fig:f18}
\end{figure*}

The XPSP standard stars are derived from the \textit{corrected} Gaia XP spectra, while the SCR standard stars, based on atmospheric parameters from LAMOST and Gaia photometry, are predicted with an independent approach. In the absence of an ``absolute reference", the consistency of the re-calibration zero-points obtained from these two types of standard stars provides a reasonable estimate of the zero-point precision for the re-calibrated USS DR1 data.

A comparison of zero-points between the XPSP and the SCR standards for all bands is shown in Figure\,\ref{Fig:f14}. We can see that all the points are consistently distributed along the $y=x$ line for each filter. Also, the bluer the filter, the larger the scatter. To quantitatively estimate the consistency, Figure\,\ref{Fig:f15} displays the zero-point differences between the XPSP and SCR standards, as a function of the numbers of standard stars in each tile. For each band, the standard deviations initially 
exhibit higher values, then quickly decrease and slowly converge to a stable value as the star numbers increase. The stable value, about 4--6\,mmag for the blue filters and roughly 1--2\,mmag for the others (as shown in the second line of Table\,\ref{tab:pre}) represents the final zero-point precision of the re-calibrated USS DR1 photometry.

\section{Discussion}\label{sec:dis}
\subsection{Validation of the Re-calibrated USS Photometry with SDSS Stripe 82 and PS1 Photometry}
\begin{figure*}[ht!] \centering
\subfigure{\includegraphics[width=9cm]{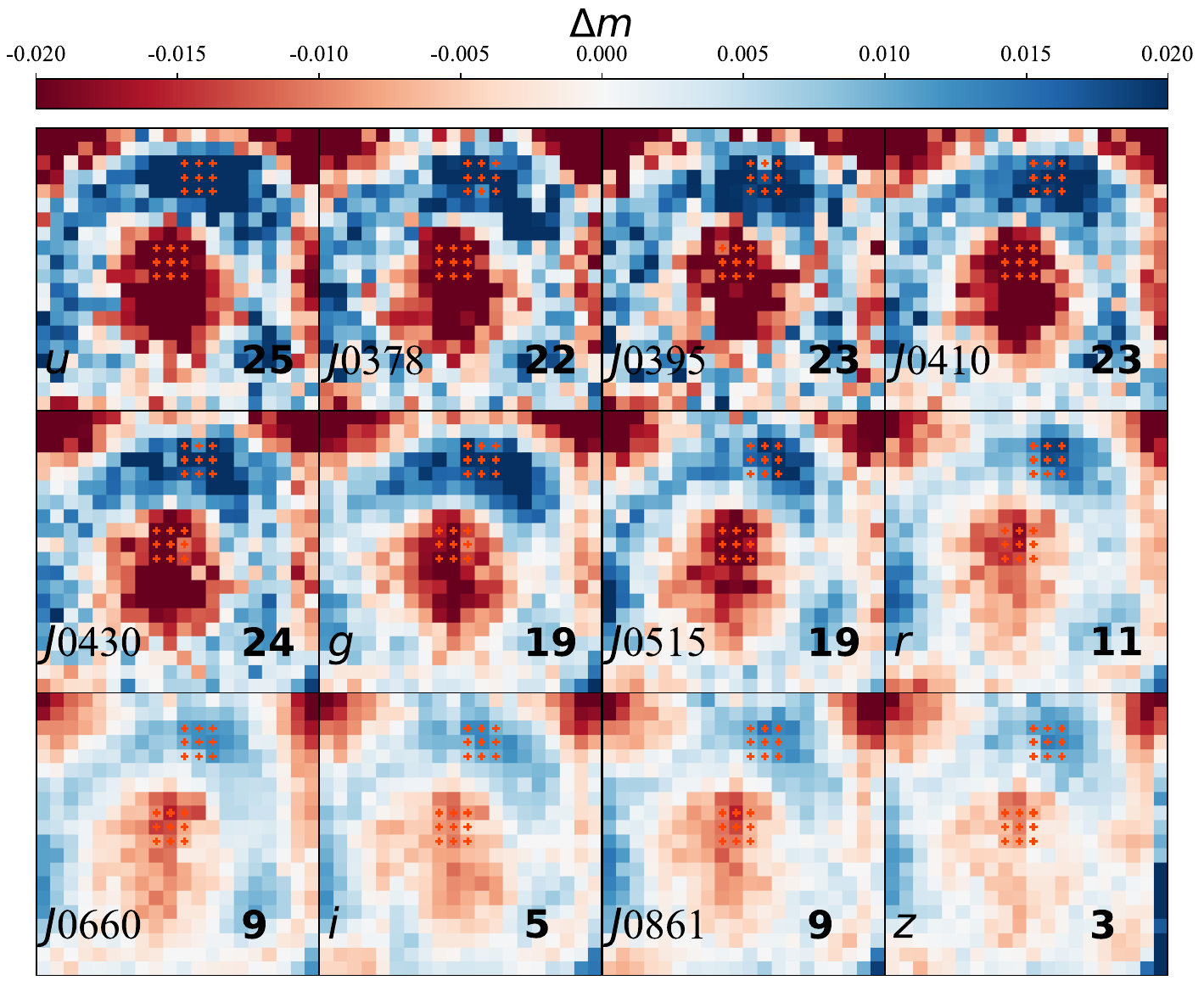}
           \includegraphics[width=9cm]{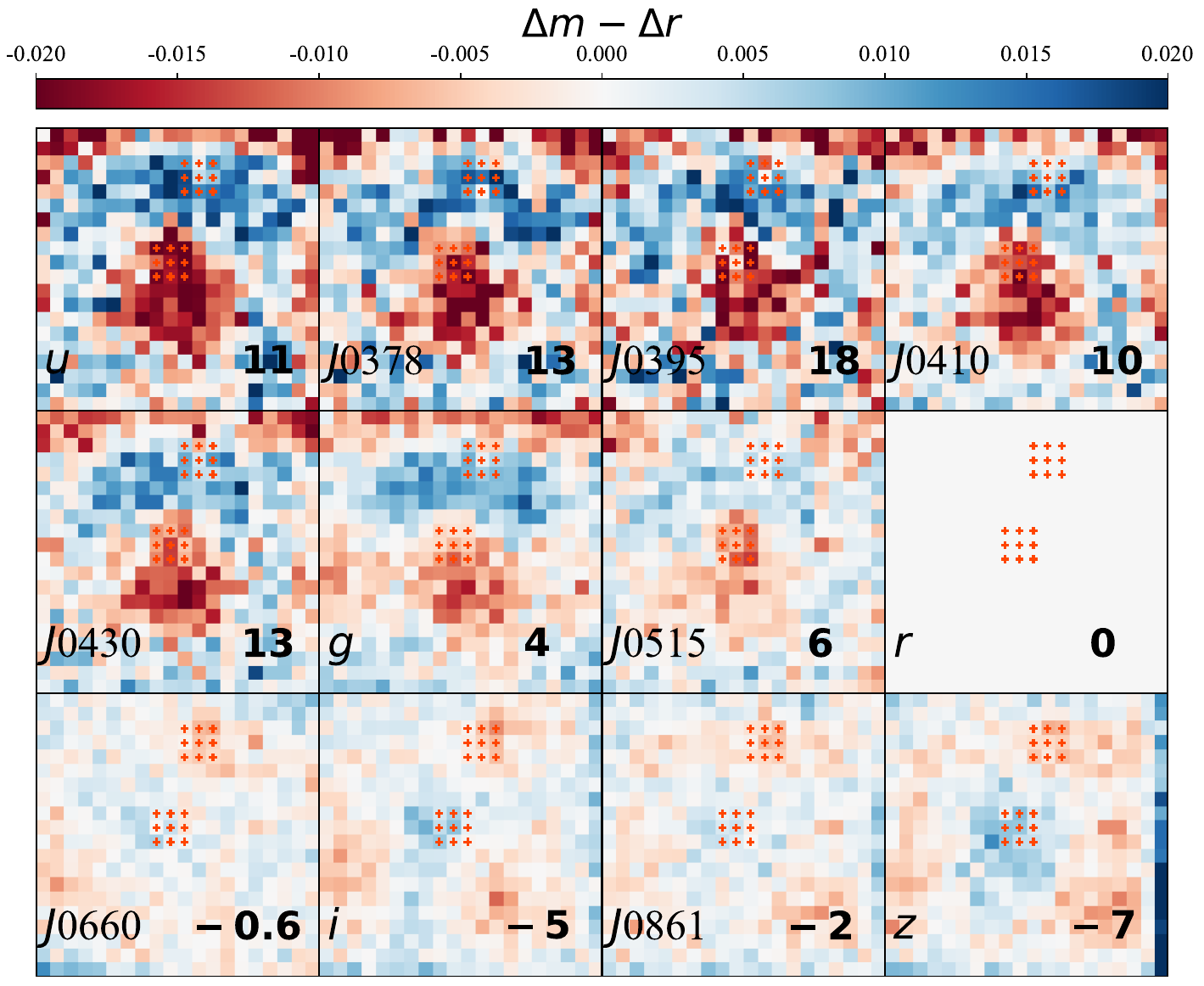}} \\
\caption{{\small The left panels show the spatial variations of the medium-scale flat-field after $20\times 20$ binning for all bands. The right panels are similar to the left panels, but display the difference between the results of all bands and the $r$-band. The bands and color bars are shown at the bottom-left corner of each panel and at the top, respectively. Nine adjacent points from the annular and central regions for each band are selected and marked with red plus-signs in each panel. The difference between the median value of the two regions, in unit of milli-magnitude, are calculated and indicated in the bottom-right corner of each panel.}}
\label{Fig:f21}
\end{figure*}

\begin{figure*}[ht!] \centering
\resizebox{\hsize}{!}{\includegraphics{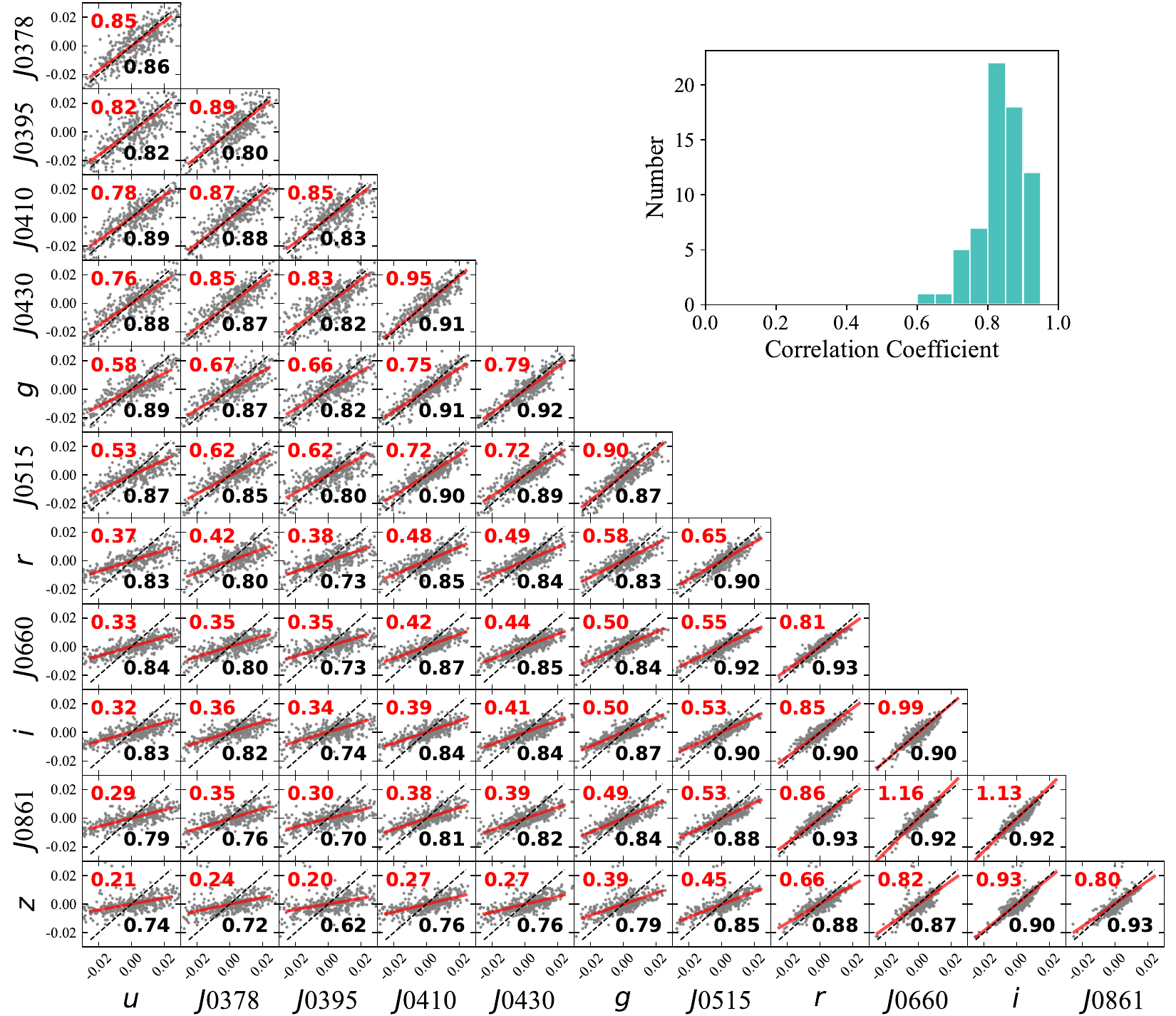}} \\
\caption{{\small Same as Figure\,\ref{Fig:f6}, but for the correlation of the 400 values for each of the two bands. For each panel, the correlation coefficients are marked in black in the bottom-right corners. The linear fitting lines are shown as red lines, and the slopes of the line are marked in in red in the top-left corners. The black-dashed lines denote $y=x$ in each panel. A histogram distribution of the correlation coefficients is shown in the top-right corner.}}
\label{Fig:f22}
\end{figure*}

To evaluate the effectiveness of the photometric re-calibration, we independently validate the re-calibrated USS photometry using the ``corrected'' PS1 DR1 photometry \citep[\url{https://nadc.china-vo.org/bestphot}, please refer to][]{2022AJ....163..185X, 2023ApJS..268...53X} and the ``corrected'' SDSS Stripe 82 standard stars catalog \citep[v2.6; \url{https://faculty.washington.edu/ivezic/sdss/catalogs/stripe82.html}, see][]{2022ApJS..259...26H}, focusing on the $r$- and $i$-bands. Figure\,\ref{Fig:f16} shows the normalized transmission functions of the $r$- and $i$-band filters for the USS, PS1, and SDSS systems, as a function of wavelength. The USS $r$- and $i$-bands are very similar to those of the SDSS and PS1 bands.

We combine the re-calibrated USS DR1 photometric data with the PS1 DR1 and SDSS Stripe 82 standards catalog with a cross-matching radius of $1^{\prime\prime}$. Then we select stars with USS $r$-band magnitudes greater than 10 and less than 14, as well as with \texttt{SEX\_FLAGS} $=$ 0, resulting a total of 723 stars. For each band, four colors: $m_{\rm SDSS}-m^{\rm corr}$, $m_{\rm PS1}-m^{\rm corr}$, $m_{\rm SDSS}-m$, and $m_{\rm PS1}-m$, are adopted. Here, $m_{\rm SDSS}$, $m_{\rm PS1}$, $m$, and $m^{\rm corr}$ represent the $m$-band magnitudes of SDSS Stripe 82, PS1, and the USS DR1 magnitude before and after re-calibration, respectively. Next, a linear polynomial, as a function of $g_{\rm SDSS}-i_{\rm SDSS}$ and 
$g_{\rm PS1}-i_{\rm PS1}$, is used to fit the colors $m_{\rm SDSS}-m$ or $m_{\rm SDSS}-m^{\rm corr}$, and $m_{\rm PS1}-m$ or $m_{\rm PS1}-m^{\rm corr}$, respectively.

The final fitting results for the $r$-band is presented in Figure\,\ref{Fig:f17}. Before photometric re-calibration, the fitting residuals are 31\,mmag for SDSS Stripe 82 and 30\,mmag for PS1; after re-calibration, they significantly decrease to about 15\,mmag, indicating a two-fold precision improvement in the re-calibration process. A similar phenomenon is also observed for the $i$-band. We ignored the effect of extinction in this process, as the extinction values for USS stars are small ($E(B-V)\le 0.068$), and the differences of the $r$- or $i$-band filter between USS, PS1 and SDSS are minimal, resulting in a low reddening coefficient \citep[less than 0.02; estimated from][]{2023ApJS..264...14Z, 2024ApJS..271...41X} in the $r_{\rm SDSS}-r$, $r_{\rm PS1}-r$, $i_{\rm SDSS}-i$, and $i_{\rm PS1}-i$ colors.

\subsection{Validation of the Re-calibrated USS Photometry with LAMOST and Gaia Photometry}
In this subsection, we describe a comprehensive evaluation across all 12 bands. We combine the USS photometry with the LAMOST DR10 and Gaia EDR3 photometry corrected by \cite{2021ApJ...908L..24Y}, using a cross-matching radius of $1^{\prime\prime}$. We then selected a low-extinction sample with $E(G_{\rm BP}-G_{\rm RP})\le 0.02$, applying the same criteria to the SCR standard stars and the calibration stars. Ultimately, a total number of 316, 382, 315, 309, 332, 368, 319, 292, 387, 298, 329 and 277 stars were selected for the $u$-, $J0378$-, $J0395$-, $J0410$-, $J0430$-, $g$-, $J0515$-, $r$-, $J0660$-, $i$, $J0861$-, and $z$-bands, respectively.

For each USS magnitude, both before and after photometric re-calibration, as well as for the XPSP standard photometry, twelve intrinsic colors were determined, following \cite{2024ApJS..271...41X}. A second-order two-dimensional polynomial, as a function of $T_{\rm eff}$ and $\rm [Fe/H]$, was then employed to fit the intrinsic colors. 

The final fitting results are presented in Figure\,\ref{Fig:f18}. Before photometric re-calibration, the standard deviation of the fitting residuals are 68, 63, 64, 50, 47, 36, 38, 32, 35, 41, 30, and 26\,mmag for the $G_{\rm BP}-u$, $G_{\rm BP}-J0378$, $G_{\rm BP}-J0395$, $G_{\rm BP}-J0410$, $G_{\rm BP}-J0430$, $G_{\rm BP}-J0515$, $G_{\rm RP}-J0660$, $G_{\rm RP}-J0861$, $G_{\rm BP}-g$, $G_{\rm RP}-r$, $G_{\rm RP}-i$, and $G_{\rm RP}-z$ colors, respectively. 
After re-calibration, the standard deviation of the fitting residuals significantly improves, decreasing to approximately 16--26\,mmag for the blue filters and 4--15\,mmag for the other filters, representing a 2- to 6-fold improvement, respectively. 

\subsection{Precision Comparison Between Re-calibrated USS DR1 Photometry and XPSP Standard Stars}
The re-calibrated USS DR1 photometry not only enables research that requires high precision, such as metallicity measurements based on photometry \citep[see, e.g.,][]{2024arXiv240802171H}, but also offers the possibility for precise photometric calibration of future USS observations. Thus, we are particularly interested in comparing the precision of the re-calibrated USS data with that of XPSP to ascertain which is superior.

The fitting results of the intrinsic colors, derived from Gaia photometry and the XPSP standards in the USS system, as a function of $T_{\rm eff}$ and $\rm [Fe/H]$, are shown in the bottom panel of Figure\,\ref{Fig:f18}. The standard deviation of the fitting residuals is approximately 10--30\,mmag for the blue filters ($u$, $J0378$, $J0395$, $J0410$, $J0430$) and 1--6\,mmag for the other filters, based on the Gaia photometry and LAMOST spectroscopic data. Our results show that the precision of individual stars for the re-calibrated data in the $u$-, $J0395$- and $z$-bands is comparable to that of the XPSP standard stars, while the precision in the other bands is significantly lower.

\subsection{The Medium-scale Flat-field Structure}\label{mottling}

This section provides a detailed description of the wavelength dependence and the correlation across different bands of the medium-scale flat-field structure. The medium-scale flat-field structure, representing the CCD spatial distribution of the difference between the XPSP and USS magnitudes after second-order polynomial correction, is characterized by 400 points per band, with the CCD space divided into $20\times 20$ bins and the median value used for each bin. 

The left panels of Figure\,\ref{Fig:f21} show the distribution of the medium-scale flat-field structure in the CCD space across 12 bands. This structure is more pronounced at the blue filters, having shorter wavelengths, and weakens as the wavelength increases. To quantify this, we selected 9 adjacent points from the annular and central regions for each band, and calculated the median difference between these two regions. The median differences are 25, 22, 23, 23, 24, 19, 19, 11, 9, 5, 9 and 3\,mmag for the $u$-, $J0378$-, $J0395$-, $J0410$-, $J0430$-, $g$-, $J0515$-, $r$-, $J0660$-, $i$-, $J0861$-, and $z$-bands, respectively. To better illustrate this dependence, we subtracted the medium-scale structure of the $r$-band from that of the other 12 bands, as shown in the right panel of Figure\,\ref{Fig:f21}. The median differences between two regions, after subtracting the $r$-band structure, are 11, 13, 18, 10, 13, 4, 6, 0, $-0.6$, $-5$, $-2$ and $-7$\,mmag for the $u$-, $J0378$-, $J0395$-, $J0410$-, $J0430$-, $g$-, 
$J0515$-, $r$-, $J0660$-, $i$-, $J0861$-, and $z$-bands, respectively.

To investigate the correlation of the medium-scale structure across different bands, we plotted the correlation of 400 points between various band pairs, as shown in Figure\,\ref{Fig:f22}. An obvious correlation is observed for each band pair. We then calculated the linear correlation coefficients; their histogram distribution is also displayed in the top-right corner of Figure\,\ref{Fig:f22}. Approximately 79\% of the band pairs have a linear correlation coefficient greater than 0.8. Furthermore, band pairs with adjacent central wavelengths exhibit points that are nearly symmetrically distributed around the $y=x$ line, while band pairs with more distant central wavelengths shown some inclination. To quantify this relationship, we performed linear fitting for each band pair; the fitting results are also presented in Figure\,\ref{Fig:f22}. The closer the central wavelengths of the band pairs, the closer the fitting slope is to 1. Otherwise, it approaches 0. The slope indicates the relative strength of the medium-scale structure between the bands in each band pair.

\section{Summary and Conclusions} \label{sec:conclusion}
In this work, we independently validated the USS DR1 photometric calibration using 3,000 to 70,000 carefully selected standard stars from the BEST database, including SCR and XPSP standard stars. 

By comparing BEST stars with USS photometry, we identified spatially dependent systematic errors of about 30--40\,mmag in the blue filters ($u$, $J0378$, $J0395$), and around 10\,mmag in other filters. These errors presented a correlation across bands, likely due to improper handling of stellar metallicity in the SL method and the ATLAS catalog used for the original calibration.

We also discovered position-dependent CCD systematic errors up to 50\,mmag, which correlated with the spatial distribution of the difference between USS \texttt{APER\_6} and USS magnitudes, and anti-correlated with the distribution of FWHM. We developed a linear model to correct these errors based on FWHM and used a second-order polynomial in $X$ and $Y$ to correct the remaining error. After the corrections, we identified and corrected medium-scale errors up to 50\,mmag between large-scale and small-scale flats within the CCD space.

Independent validation using SDSS Stripe 82 and corrected PS1 data presented a 5-fold decrease in scatter in the color-color diagrams after re-calibration. Validations with LAMOST DR10 and Gaia photometry confirmed a 2- to 6-fold improvement. These results highlight the effectiveness of the re-calibration and the utility of the BEST database.

We recommend that future USS observations use the BEST database for high-precision photometric calibration, as demonstrated in this study.

\clearpage

We sincerely thank the S-PLUS collaboration for providing the USS DR1 photometric data.
This work is supported by the National Key Basic R\&D Program of China via 2023YFA1608300, the NSFC grant No. 12403024, the National Natural Science Foundation of China through the projects NSFC 12222301, 12173007, the Postdoctoral Fellowship Program of CPSF under Grant Number GZB20240731, the Young Data Scientist Project of the National Astronomical Data Center, and the China Post-doctoral Science Foundation No. 2023M743447.
Y.K.T. is supported by the International Centre of Supernovae, Yunnan Key Laboratory (Nos. 202302AN360001 and 202302AN36000103).
T.C.B. acknowledges partial support from grant PHY 14-30152; Physics Frontier Center/JINA Center for the Evolution of the Elements (JINA-CEE), and from OISE-1927130: The International Research Network for Nuclear Astrophysics (IReNA), awarded by the US National Science Foundation. 
P.K.H gratefully acknowledges the Fundação de Amparo à Pesquisa do Estado de São Paulo (FAPESP) for the support grant 2023/14272-4.
A.A.C acknowledges financial support from the Severo Ochoa grant CEX2021-001131-S funded by MCIN/AEI/10.13039/501100011033 and the Spanish project PID2023-153123NB-I00 funded by MCIN/AEI.

The S-PLUS project, including the T80-South robotic telescope and the S-PLUS scientific survey, was founded as a partnership between the Fundação de Amparo à Pesquisa do Estado de São Paulo (FAPESP), the Observatório Nacional (ON), the Federal University of Sergipe (UFS), and the Federal University of Santa Catarina (UFSC), with important financial and practical contributions from other collaborating institutes in Brazil, Chile (Universidad de La Serena), and Spain (Centro de Estudios de Física del Cosmos de Aragón, CEFCA). We further acknowledge financial support from the São Paulo Research Foundation (FAPESP), Fundação de Amparo à Pesquisa do Estado do RS (FAPERGS), the Brazilian National Research Council (CNPq), the Coordination for the Improvement of Higher Education Personnel (CAPES), the Carlos Chagas Filho Rio de Janeiro State Research Foundation (FAPERJ), and the Brazilian Innovation Agency (FINEP). The authors who are members of the S-PLUS collaboration are grateful for the contributions from CTIO staff in helping in the construction, commissioning and maintenance of the T80-South telescope and camera. 

We are also indebted to Rene Laporte and INPE, as well as Keith Taylor, for their important contributions to the project. From CEFCA, we particularly would like to thank Antonio Marín-Franch for his invaluable contributions in the early phases of the project, David Cristóbal-Hornillos and his team for their help with the installation of the data reduction package jype version 0.9.9, César Íñiguez for providing 2D measurements of the filter transmissions, and all other staff members for their support with various aspects of the project.

This work has made use of data from the European Space Agency (ESA) mission
Gaia (\url{https://www.cosmos.esa.int/gaia}), processed by the Gaia
Data Processing and Analysis Consortium (DPAC,
\url{https://www.cosmos.esa.int/web/gaia/dpac/consortium}). Funding for the DPAC
has been provided by national institutions, in particular the institutions
participating in the Gaia Multilateral Agreement.
Guoshoujing Telescope (the Large Sky Area Multi-Object Fiber Spectroscopic Telescope LAMOST) is a National Major Scientific Project built by the Chinese Academy of Sciences. Funding for the project has been provided by the National Development and Reform Commission. LAMOST is operated and managed by the National Astronomical Observatories, Chinese Academy of Sciences.

\software{Astropy \citep{2022ApJ...935..167A}, Matplotlib \citep{2007CSE.....9...90H}, NumPy \citep{2020Natur.585..357H}, SciPy \citep{2020NatMe..17..261V}}

\clearpage
\appendix
\setcounter{table}{0}   
\setcounter{figure}{0}
\renewcommand{\thetable}{A\arabic{table}}
\renewcommand{\thefigure}{A\arabic{figure}}

\begin{figure}[ht!] \centering
\includegraphics[width=15cm]{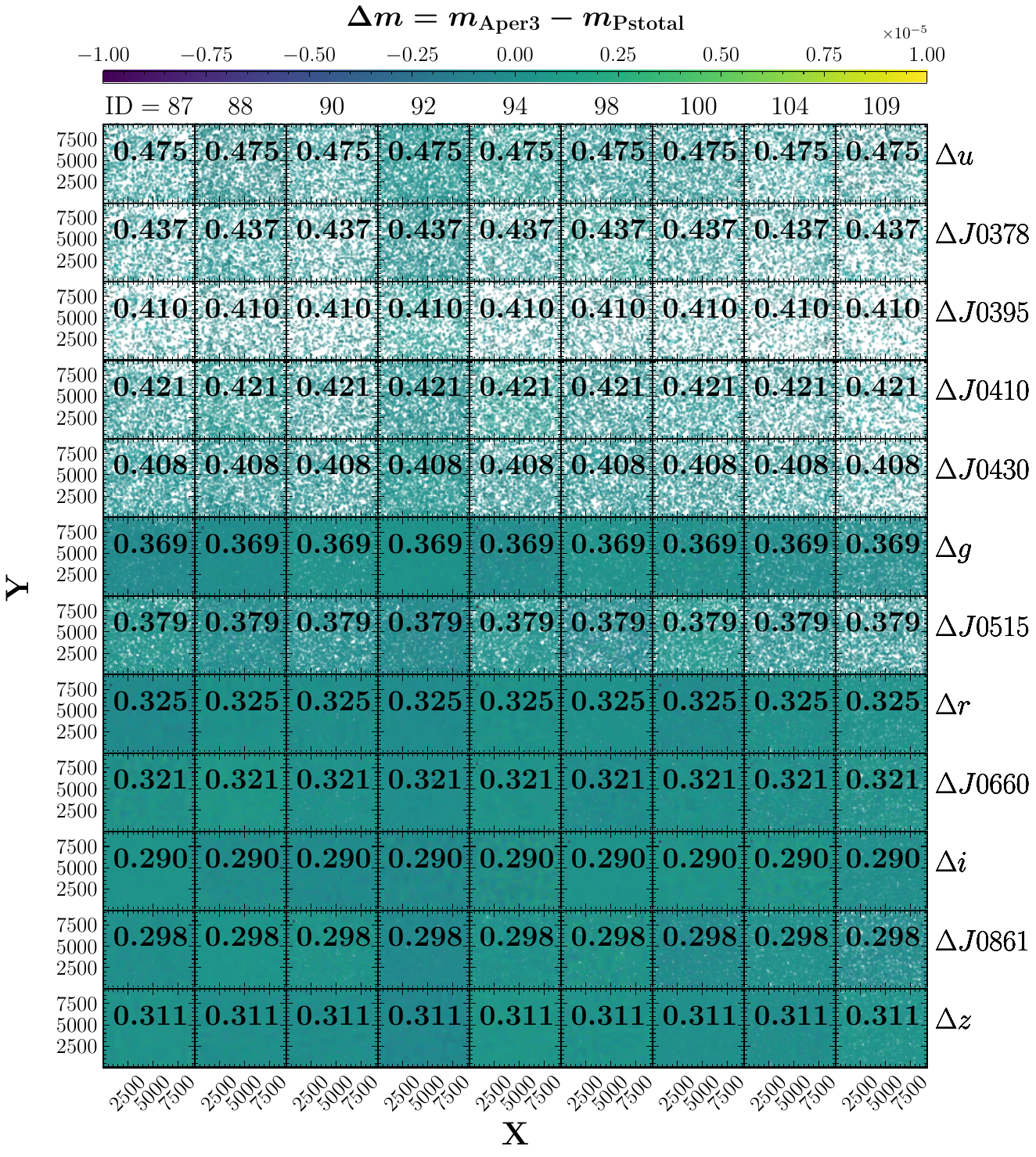}
\caption{{\small An example showing the CCD spatial distribution of the magnitude difference between \texttt{Aper\_3} and \texttt{Pstotal} for all observed stars within nine tiles.  The top to bottom panels shows the results of the $u$-, $J0378$-, $J0395$-, $J0410$-, $J0430$-, $g$-, $J0515$-, $r$-, $J0660$-, $i$-, $J0861$-, and $z$-bands, respectively, with band labels at the 
right and a color bar on the top. The selected nine tiles (\texttt{SHORTS-STRIPE82\_0087}, \texttt{SHORTS-STRIPE82\_0088}, \texttt{SHORTS-STRIPE82\_0090}, \texttt{SHORTS-STRIPE82\_0092}, \texttt{SHORTS-STRIPE82\_0094}, \texttt{SHORTS-STRIPE82\_0098}, \texttt{SHORTS-STRIPE82\_0100}, \texttt{SHORTS-STRIPE82\_0104}, and \texttt{SHORTS-STRIPE82\_0109}) are ordered from left to right, with the tile number marked between the first-row panels and the color bar. The median value of the magnitude difference between \texttt{Aper\_3} and \texttt{Pstotal} is marked for each panel. Note that the color bar ranges from $-1\times 10^{-5}$ to $1\times 10^{-5}$.}}
\label{Fig:A0}
\end{figure}

\begin{figure}[ht!] \centering
\resizebox{\hsize}{!}{\includegraphics{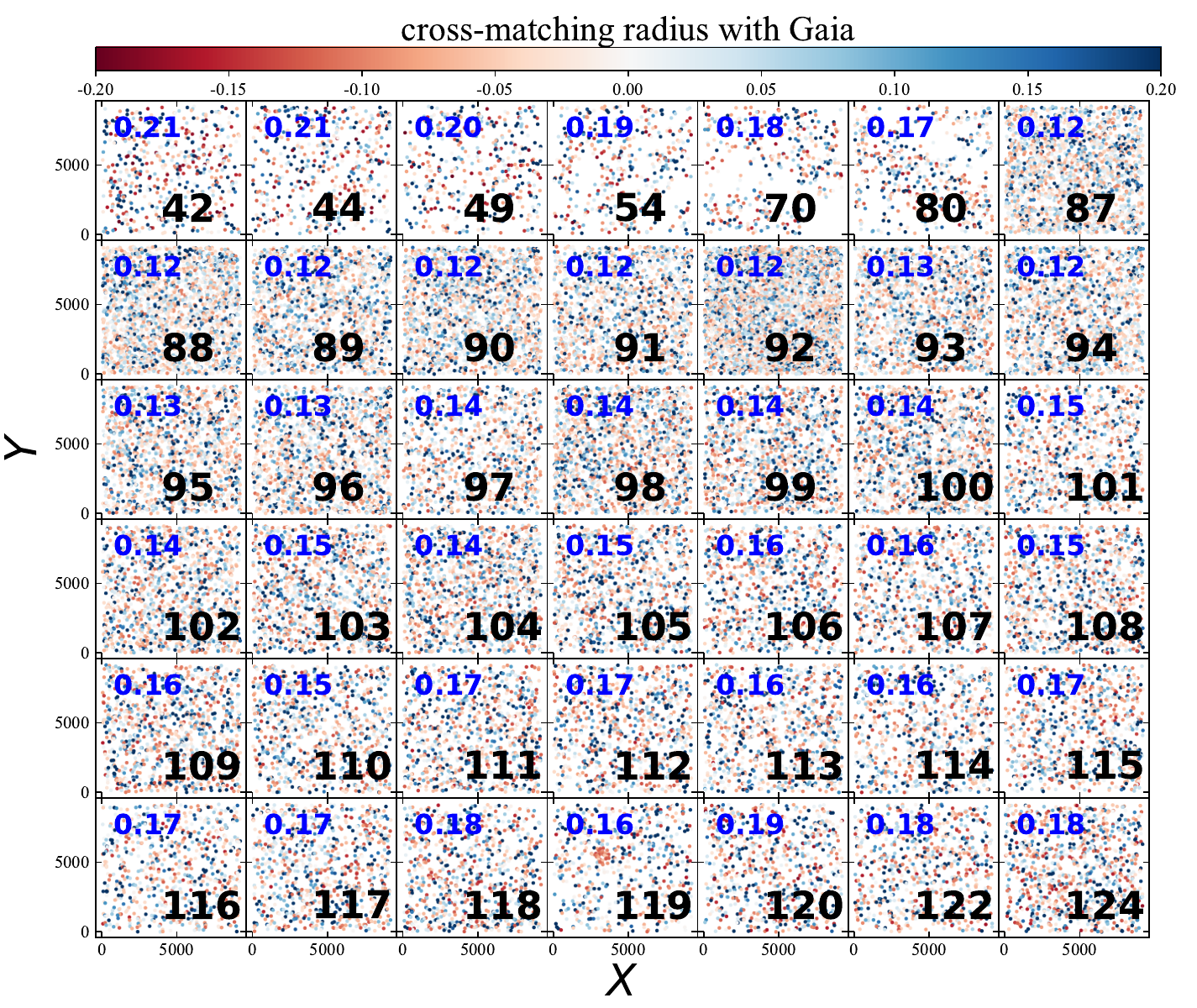}}
\caption{{\small Same as Figure\,\ref{Fig:f9}, but for the cross-matching radius with Gaia. To clearly display the spatial distribution of the astrometric offsets between USS and Gaia, we calculated the median value of astrometric offset for each tile. We then subtracted this median value from all offsets of stars within the tile, and the resulting astrometric offsets are shown in each panel. The \texttt{tile\_ID} and the median value of astrometric offset are labeled in the top-left and the bottom-right corner of each panel, respectively.}}
\label{Fig:A2}
\end{figure}

\begin{figure}[ht!] \centering
\resizebox{\hsize}{!}{\includegraphics{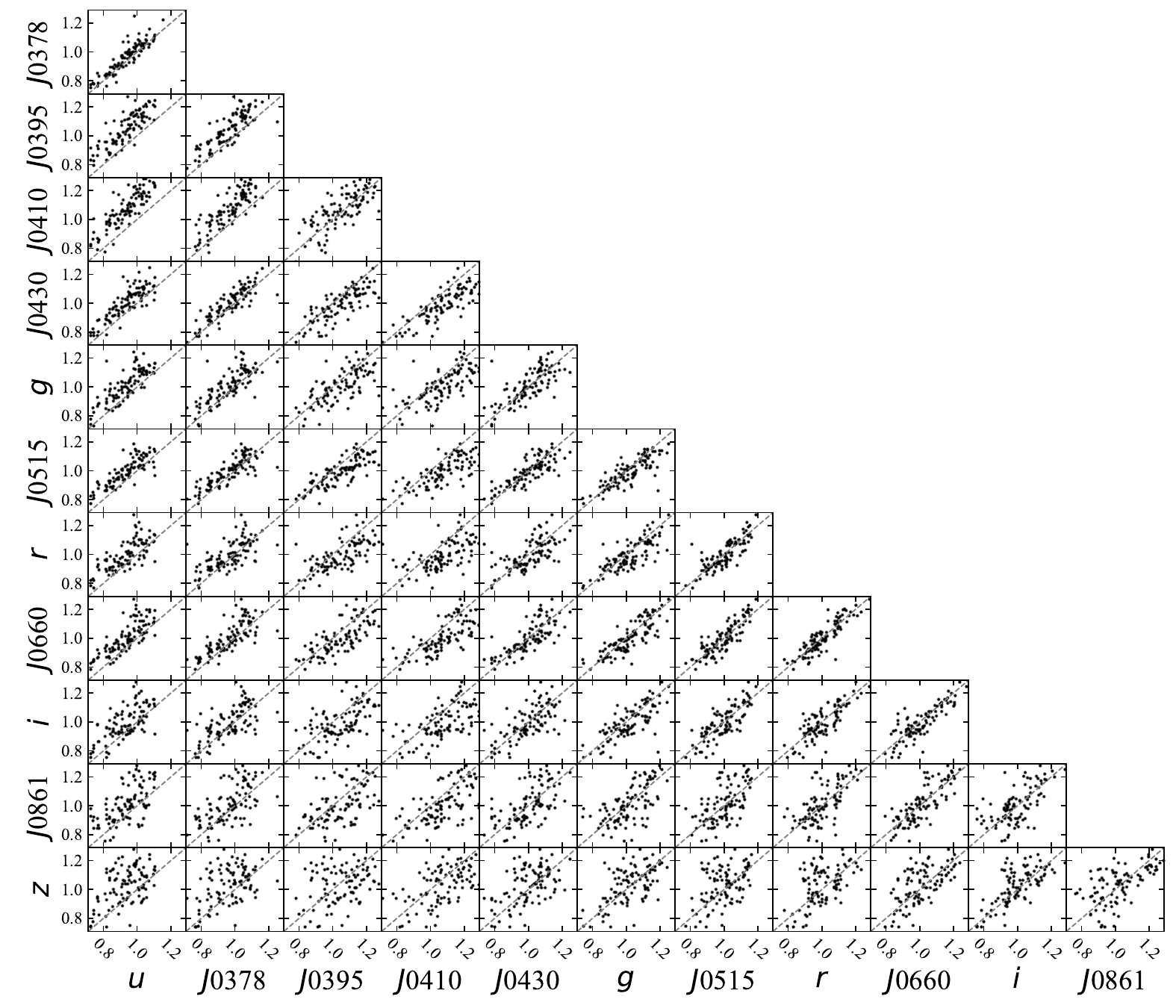}}
\caption{{\small Correlation plots of the normalized FWHM for each of the two bands for \texttt{SHORTS-STRIPE82\_0090}. The 
gray-dashed lines denote $y=x$ in each panel.}}
\label{Fig:A3}
\end{figure}

This appendix provides the spatial distribution of the magnitude difference between \texttt{Aper\_3} and \texttt{Pstotal}, and the FWHM correlation for each band pair in the \texttt{SHORTS-STRIPE82\_0090} tile,  shown in Figure\,\ref{Fig:A0} and Figure\,\ref{Fig:A3}, respectively.  Examples of the spatial variations of the astrometric offsets between Gaia and USS in the $r$-band are shown in Figure\,\ref{Fig:A2}.

\end{CJK}

\clearpage

\begin{thebibliography}{} \label{bib}
\bibitem[Astropy Collaboration et al.(2022)]{2022ApJ...935..167A} Astropy Collaboration, Price-Whelan, A.~M., Lim, P.~L., et al.\ 2022, \apj, 935, 167. doi:10.3847/1538-4357/ac7c74
\bibitem[Almeida-Fernandes et al.(2022)]{2022MNRAS.511.4590A} Almeida-Fernandes, F., SamPedro, L., Herpich, F.~R., et al.\ 2022, \mnras, 511, 4590. doi:10.1093/mnras/stac284
\bibitem[Benitez et al.(2014)]{2014arXiv1403.5237B} Benitez, N., Dupke, R., Moles, M., et al.\ 2014, arXiv:1403.5237. doi:10.48550/arXiv.1403.5237
\bibitem[Bertin \& Arnouts(1996)]{1996A&AS..117..393B} Bertin, E. \& Arnouts, S.\ 1996, \aaps, 117, 393. doi:10.1051/aas:1996164
\bibitem[Carrasco et al.(2021)]{2021A&A...652A..86C} Carrasco, J.~M., Weiler, M., Jordi, C., et al.\ 2021, \aap, 652, A86. doi:10.1051/0004-6361/202141249
\bibitem[Cenarro et al.(2019)]{2019A&A...622A.176C} Cenarro, A.~J., Moles, M., Crist{\'o}bal-Hornillos, D., et al.\ 2019, \aap, 622, A176. doi:10.1051/0004-6361/201833036
\bibitem[Clem \& Landolt(2013)]{2013AJ....146...88C} Clem, J.~L. \& Landolt, A.~U.\ 2013, \aj, 146, 88. doi:10.1088/0004-6256/146/4/88
\bibitem[Clem \& Landolt(2016)]{2016AJ....152...91C} Clem, J.~L. \& Landolt, A.~U.\ 2016, \aj, 152, 91. doi:10.3847/0004-6256/152/4/91
\bibitem[Cui et al.(2012)]{2012RAA....12.1197C} Cui, X.-Q., Zhao, Y.-H., Chu, Y.-Q., et al.\ 2012, Research in Astronomy and Astrophysics, 12, 1197. doi:10.1088/1674-4527/12/9/003
\bibitem[De Angeli et al.(2023)]{2023A&A...674A...2D} De Angeli, F., Weiler, M., Montegriffo, P., et al.\ 2023, \aap, 674, A2. doi:10.1051/0004-6361/202243680
\bibitem[De Silva et al.(2015)]{2015MNRAS.449.2604D} De Silva, G.~M., Freeman, K.~C., Bland-Hawthorn, J., et al.\ 2015, \mnras, 449, 2604. doi:10.1093/mnras/stv327
\bibitem[Deng et al.(2012)]{2012RAA....12..735D} Deng, L.-C., Newberg, H.~J., Liu, C., et al.\ 2012, Research in Astronomy and Astrophysics, 12, 735. doi:10.1088/1674-4527/12/7/003
\bibitem[Fan et al.(2023)]{2023ApJS..268....9F} Fan, Z., Zhao, G., Wang, W., et al.\ 2023, \apjs, 268, 9. doi:10.3847/1538-4365/ace04a
\bibitem[Gaia Collaboration et al.(2021a)]{2021A&A...649A...1G} Gaia Collaboration, Brown, A.~G.~A., Vallenari, A., et al.\ 2021a, \aap, 649, A1. doi:10.1051/0004-6361/202039657
\bibitem[Gaia Collaboration et al.(2021b)]{2021A&A...650C...3G} Gaia Collaboration, Brown, A.~G.~A., Vallenari, A., et al.\ 2021b, \aap, 650, C3. doi:10.1051/0004-6361/202039657e
\bibitem[Gaia Collaboration et al.(2018)]{2018A&A...616A...1G} Gaia Collaboration, Brown, A.~G.~A., Vallenari, A., et al.\ 2018, \aap, 616, A1. doi:10.1051/0004-6361/201833051
\bibitem[Gaia Collaboration et al.(2023a)]{2023A&A...674A...1G} Gaia Collaboration, Vallenari, A., Brown, A.~G.~A., et al.\ 2023a, \aap, 674, A1. doi:10.1051/0004-6361/202243940
\bibitem[Gaia Collaboration et al.(2023b)]{2023A&A...674A..33G} Gaia Collaboration, Montegriffo, P., Bellazzini, M., et al.\ 2023, \aap, 674, A33. doi:10.1051/0004-6361/202243709
\bibitem[Harris et al.(2020)]{2020Natur.585..357H} Harris, C.~R., Millman, K.~J., van der Walt, S.~J., et al.\ 2020, \nat, 585, 357. doi:10.1038/s41586-020-2649-2
\bibitem[Huang \& Yuan(2022)]{2022ApJS..259...26H} Huang, B. \& Yuan, H.\ 2022, \apjs, 259, 26. doi:10.3847/1538-4365/ac470d
\bibitem[Huang et al.(2022)]{2022SSPMA..52B9503H} Huang, B., Xiao, K., \& Yuan, H.\ 2022, Scientia Sinica Physica, Mechanica \& Astronomica, 52, 289503. doi:10.1360/SSPMA-2022-0086
\bibitem[Huang et al.(2024a)]{2024ApJS..271...13H} Huang, B., Yuan, H., Xiang, M., et al.\ 2024a, \apjs, 271, 13. doi:10.3847/1538-4365/ad18b1
\bibitem[Huang et al.(2021)]{2021ApJ...907...68H} Huang, Y., Yuan, H., Li, C., et al.\ 2021, \apj, 907, 68. doi:10.3847/1538-4357/abca37
\bibitem[Huang et al.(2024)]{2024arXiv240802171H} Huang, Y., Beers, T.~C., Xiao, K., et al.\ 2024, arXiv:2408.02171. doi:10.48550/arXiv.2408.02171
\bibitem[Hunter(2007)]{2007CSE.....9...90H} Hunter, J.~D.\ 2007, Computing in Science and Engineering, 9, 90. doi:10.1109/MCSE.2007.55
\bibitem[Ivezi{\'c} et al.(2007)]{2007AJ....134..973I} Ivezi{\'c}, {\v{Z}}., Smith, J.~A., Miknaitis, G., et al.\ 2007, \aj, 134, 973. doi:10.1086/519976
\bibitem[Ivezi{\'c} et al.(2019)]{2019ApJ...873..111I} Ivezi{\'c}, {\v{Z}}., Kahn, S.~M., Tyson, J.~A., et al.\ 2019, \apj, 873, 111. doi:10.3847/1538-4357/ab042c
\bibitem[Liu et al.(2014)]{2014IAUS..298..310L} Liu, X.-W., Yuan, H.-B., Huo, Z.-Y., et al.\ 2014, Setting the scene for Gaia and LAMOST, 298, 310. doi:10.1017/S1743921313006510
\bibitem[Liu et al.(2021)]{2021AnABC..93..628L} Liu, J., Soria, R., Wu, X.-F., et al.\ 2021, Anais da Academia Brasileira de Ciencias, 93, 20200628. doi:10.1590/0001-3765202120200628
\bibitem[L{\'o}pez-Sanjuan et al.(2019)]{2019A&A...631A.119L} L{\'o}pez-Sanjuan, C., Varela, J., Crist{\'o}bal-Hornillos, D., et al.\ 2019, \aap, 631, A119. doi:10.1051/0004-6361/201936405
\bibitem[L{\'o}pez-Sanjuan et al.(2024)]{2024A&A...683A..29L} L{\'o}pez-Sanjuan, C., V{\'a}zquez Rami{\'o}, H., Xiao, K., et al.\ 2024, \aap, 683, A29. doi:10.1051/0004-6361/202346012
\bibitem[Mendes de Oliveira et al.(2019)]{2019MNRAS.489..241M} Mendes de Oliveira, C., Ribeiro, T., Schoenell, W., et al.\ 2019, \mnras, 489, 241. doi:10.1093/mnras/stz1985
\bibitem[Montegriffo et al.(2023)]{2023A&A...674A...3M} Montegriffo, P., De Angeli, F., Andrae, R., et al.\ 2023, \aap, 674, A3. doi:10.1051/0004-6361/202243880
\bibitem[Niu et al.(2021a)]{2021ApJ...909...48N} Niu, Z., Yuan, H., \& Liu, J.\ 2021a, \apj, 909, 48. doi:10.3847/1538-4357/abdbac
\bibitem[Niu et al.(2021b)]{2021ApJ...908L..14N} Niu, Z., Yuan, H., \& Liu, J.\ 2021b, \apjl, 908, L14. doi:10.3847/2041-8213/abe1c2
\bibitem[Perottoni et al.(2024)]{2024arXiv240705004P} Perottoni, H.~D., Placco, V.~M., Almeida-Fernandes, F., et al.\ 2024, arXiv:2407.05004. doi:10.48550/arXiv.2407.05004
\bibitem[Ruz-Mieres(2022)]{2022zndo...6674521R} Ruz-Mieres, D.\ 2022, Zenodo
\bibitem[S{\'a}nchez-Bl{\'a}zquez et al.(2006)]{miles} S{\'a}nchez-Bl{\'a}zquez, P., Peletier, R.~F., Jim{\'e}nez-Vicente, J., et al.\ 2006, \mnras, 371, 703. doi:10.1111/j.1365-2966.2006.10699.x
\bibitem[Schlegel et al.(1998)]{1998ApJ...500..525S} Schlegel, D.~J., Finkbeiner, D.~P., \& Davis, M.\ 1998, \apj, 500, 525. doi:10.1086/305772
\bibitem[Stubbs \& Tonry(2006)]{2006ApJ...646.1436S} Stubbs, C.~W. \& Tonry, J.~L.\ 2006, \apj, 646, 1436. doi:10.1086/505138
\bibitem[Sun et al.(2022)]{2022ApJS..260...17S} Sun, Y., Yuan, H., \& Chen, B.\ 2022, \apjs, 260, 17. doi:10.3847/1538-4365/ac642f
\bibitem[Tonry et al.(2012)]{2012ApJ...750...99T} Tonry, J.~L., Stubbs, C.~W., Lykke, K.~R., et al.\ 2012, \apj, 750, 99. doi:10.1088/0004-637X/750/2/99
\bibitem[Tonry et al.(2018)]{2018ApJ...867..105T} Tonry, J.~L., Denneau, L., Flewelling, H., et al.\ 2018, \apj, 867, 105. doi:10.3847/1538-4357/aae386
\bibitem[Virtanen et al.(2020)]{2020NatMe..17..261V} Virtanen, P., Gommers, R., Oliphant, T.~E., et al.\ 2020, Nature Methods, 17, 261. doi:10.1038/s41592-019-0686-2
\bibitem[Wolf et al.(2018)]{2018PASA...35...10W} Wolf, C., Onken, C.~A., Luvaul, L.~C., et al.\ 2018, \pasa, 35, e010. doi:10.1017/pasa.2018.5
\bibitem[Xiao \& Yuan(2022)]{2022AJ....163..185X} Xiao, K. \& Yuan, H.\ 2022, \aj, 163, 185. doi:10.3847/1538-3881/ac540a
\bibitem[Xiao et al. (2023a)]{xiao} Xiao, K., Yuan, H., Huang B., et al.\ 2023a, Chinese Science Bulletin (in Chinese). doi:10.1360/TB-2023-0052
\bibitem[Xiao et al.(2023b)]{2023ApJS..268...53X} Xiao, K., Yuan, H., Huang, B., et al.\ 2023b, \apjs, 268, 53. doi:10.3847/1538-4365/acee73
\bibitem[Xiao et al.(2023c)]{2023ApJS..269...58X} Xiao, K., Yuan, H., L{\'o}pez-Sanjuan, C., et al.\ 2023c, \apjs, 269, 58. doi:10.3847/1538-4365/ad0645
\bibitem[Xiao et al.(2024a)]{2024ApJS..271...41X} Xiao, K., Huang, Y., Yuan, H., et al.\ 2024a, \apjs, 271, 41. doi:10.3847/1538-4365/ad24fa
\bibitem[Xiao et al.(2024b)]{2024ApJ...968L..24X} Xiao, K., Huang, B., Huang, Y., et al.\ 2024b, \apjl, 968, L24. doi:10.3847/2041-8213/ad5205
\bibitem[Yang et al.(2021)]{2021ApJ...908L..24Y} Yang, L., Yuan, H., Zhang, R., et al.\ 2021, \apjl, 908, L24. doi:10.3847/2041-8213/abdbae
\bibitem[York et al.(2000)]{2000AJ....120.1579Y} York, D.~G., Adelman, J., Anderson, J.~E., et al.\ 2000, \aj, 120, 1579. doi:10.1086/301513
\bibitem[Yuan et al.(2013)]{2013MNRAS.430.2188Y} Yuan, H.~B., Liu, X.~W., \& Xiang, M.~S.\ 2013, \mnras, 430, 2188. doi:10.1093/mnras/stt039
\bibitem[Yuan et al.(2015a)]{2015ApJ...799..133Y} Yuan, H., Liu, X., Xiang, M., et al.\ 2015a, \apj, 799, 133. doi:10.1088/0004-637X/799/2/133
\bibitem[Zhan(2018)]{2018cosp...42E3821Z} Zhan, H.\ 2018, 42nd COSPAR Scientific Assembly, 42, E1.16-4-18
\bibitem[Zhang \& Yuan(2020)]{2020ApJ...905L..20R} Zhang, R. \& Yuan, H.\ 2020, \apjl, 905, L20. doi:10.3847/2041-8213/abccc4
\bibitem[Zhang \& Yuan(2023)]{2023ApJS..264...14Z} Zhang, R. \& Yuan, H.\ 2023, \apjs, 264, 14. doi:10.3847/1538-4365/ac9dfa
\bibitem[Zhao et al.(2012)]{2012RAA....12..723Z} Zhao, G., Zhao, Y.-H., Chu, Y.-Q., et al.\ 2012, Research in Astronomy and Astrophysics, 12, 723. doi:10.1088/1674-4527/12/7/002
\bibitem[Zheng et al.(2018)]{Zheng18} Zheng, J., Zhao, G., Wang, W., et al.\ 2018, Research in Astronomy and Astrophysics, 18. doi:10.1088/1674-4527/18/12/147
\bibitem[Zheng et al.(2019)]{Zheng19} Zheng, J., Zhao, G., Wang, W., et al.\ 2019, Research in Astronomy and Astrophysics, 19. doi:10.1088/1674-4527/19/1/3
\end{thebibliography}
\end{document}